\documentclass[12pt]{article}

\usepackage{epsf}
\usepackage{cite}

\catcode`@=11 \@addtoreset{equation}{section} \catcode`@=12

\begin{document}
\begin{titlepage}
\vfill
\begin{center}
{\Large \bf $SO(N)$ Reformulated Link Invariants from  
Topological Strings}\\[1cm] 
Pravina Borhade \footnote{E-mail: pravina@phy.iitb.ac.in}, 
P. Ramadevi\footnote{Email: ramadevi@phy.iitb.ac.in}\\
{\em Department of Physics, \\Indian Institute of Technology Bombay,\\
Mumbai 400 076, India\\[10pt]}
\end{center}
\vfill
\begin{abstract}
Large $N$ duality conjecture between $U(N)$ Chern-Simons gauge theory 
on $S^3$ and $A$-model topological string theory on the resolved conifold
was verified at the level of partition function and Wilson loop observables.
As a consequence, the conjectured form for the expectation value of the 
topological operators in $A$-model string theory led to a reformulation of link
invariants in $U(N)$ Chern-Simons theory giving new polynomial 
invariants whose integer coefficients could be given a topological 
meaning. We show that the $A$-model topological operator 
involving $SO(N)$ holonomy leads to a reformulation of 
link invariants in $SO(N)$ Chern-Simons theory. 
Surprisingly, the $SO(N)$ reformulated invariants also has a similar 
form with integer coefficients. The topological meaning of the 
integer coefficients needs to be explored from the duality conjecture 
relating $SO(N)$ Chern-Simons theory to $A$-model closed string theory 
on orientifold of the resolved conifold background. 
 
\end{abstract}
\vfill
\end{titlepage}

\section{Introduction}
Within the last one decade we have seen interesting
developments in the open string and closed string dualities.
One such open-closed string duality conjecture relates $A$-model
open topological string theory on the deformed conifold,
equivalent to Chern-Simons gauge theory on $S^3$ \cite {wittencs}, 
to the closed string theory on a resolved conifold. 

Gopakumar-Vafa \cite {gv1,gv2, gv3} showed that the free-energy expansion
of $U(N)$ Chern-Simons field theory on $S^3$ 
at large $N$  resembles $A$-model topological string theory amplitudes
on the resolved conifold. The conjecture was further tested at 
the level of observables in Chern-Simons theory. Ooguri-Vafa\cite{ov}
considered the expectation value of a topological operator 
corresponding to a simple circle (called unknot) in submanifold 
$S^3$ of the deformed conifold and showed its  form in the resolved 
conifold background. The results led to a new conjecture (usually
referred to as Ooguri-Vafa conjecture) on the form for the expectation 
value of the topological operator for any knot or link in $S^3$.  

Using group theory, Labastida-Marino \cite{lm} showed that the
expectation value of the topological operators can be rewritten
in terms of link invariants in $U(N)$ Chern-Simons field theory
on $S^3$. This enabled verification of Ooguri-Vafa conjecture
for many non-trivial knots \cite {lm, taps, laba, mari}. 
Conversely, the Ooguri-Vafa conjecture led to a reformulation of 
Chern-Simons field theory invariants for links giving new
polynomial invariants. The integer coefficients of these
new polynomial invariants have topological meaning accounting
for BPS states in the string theory. The challenge still
remains in obtaining such integers within topological string theory.

Similar duality conjectures have been attempted between 
Chern-Simons gauge theories on three-manifolds other
than $S^3$ and closed string theories. 
In ref. \cite{ramprav}, $U(N)$ Chern-Simons free-energy expansion
at large $N$ for many three-manifolds were derived 
and the  expansion resembled partition function of a closed string theory 
on a Calabi-Yau background with one kahler parameter. 
Unfortunately,  the Chern-Simons free-energy expansion 
for other three-manifolds are not equivalent to the `t Hooft 
large $N$ perturbative expansion around a classical solution \cite {thoo}.
Hence we need to extract the perturbative expansion 
around a classical solution from the free-energy to obtain
new duality conjectures. 

For orbifolds of $S^3$, which gives Lens space ${\cal L}[p,1] \equiv S^3/Z_p$,
it is believed that the Chern-Simons theory is dual to the
$A$-model closed string theory on $A_{p-1}$ 
fibred over $P^1$ Calabi-Yau background.  It was Marino \cite {marin} who
showed that the perturbative Chern-Simons theory on 
Lens space ${\cal L}[p,1]$ can be given a  matrix model description. 
Also, hermitian matrix model description of $B$-model topological
strings \cite {dijk} was shown to be equivalent to Marino's matrix model
using mirror symmetry\cite{akmv}. It is still a challenging open problem 
to look for dual closed string description corresponding to $U(N)$ Chern-Simons
theory on other three-manifolds.

The extension of these duality conjectures for other gauge 
groups like $SO(N)$ and $Sp(N)$ have also been studied.
In particular, the free-energy expansion of the Chern-Simons 
theory based on $SO/Sp$ gauge group was shown to be dual to 
$A$-model closed string theory on a orientifold of the resolved conifold 
background \cite {sinha}. Further, using the topological vertex as a
tool, Bouchard et al \cite {vinc1,vinc2} have determined unoriented closed
string amplitude and unoriented open topological string amplitudes
for a few orientifold toric geometry with or without $D$-branes. 

It will be interesting to generalise Ooguri-Vafa conjecture
by looking at the topological operator involving $SO/Sp$ holonomy
instead of the $U(N)$ holonomy. In this paper, we obtain
new reformulated polynomial invariants in terms of the framed
link polynomials in $SO(N)$ Chern-Simons theory. Similar to the $U(N)$
result, the coefficients are indeed integers and the topological
meaning in terms of the BPS invariants in string theory needs to  
be explored. Further, the reformulated invariant for 
knots in standard framing obeys the conjecture of 
Bouchard-Florea-Marino \cite {vinc2} giving the integer 
coefficients corresponding to cross-cap $c=1$ 
unoriented open-string amplitude. We generalise the conjecture
for any $r$-component framed links and have verified for few examples
of framed knots and two-component framed links.

The organisation of the paper is as follows. In
section 2, we present framed link invariants
in $SO(N)$ Chern-Simons theory. In section 3, first we recapitulate
the topological operator carrying $U(N)$ holonomy  and then
elaborate its generalization to $SO(N)$ holonomy.
Section 4 contains some explicit results of the reformulated
polynomial invariants. We present the integer coefficients
in the cross-cap $c=1$ unoriented string amplitudes for few framed knots
and links in section 5. In the concluding
section 6, we summarize the results obtained and 
pose open questions for future research. 
In appendix A, we present $SO(N)$ polynomials for 
few framed knots and framed links for some representations. 
In appendix B, the reformulated polynomial invariants for 
few non-trivial framed knots and framed links are presented. 

\section{$SO(N)$ Chern-Simons Gauge theory and Framed Link invariants}

Chern-Simons gauge theory on $S^3$ based on the gauge group $SO(N)$ 
is described by the following action:
\begin{equation}
S = {k \over 4 \pi} \int_{S^3} Tr\left (A \wedge dA + {2 \over 3} A \wedge
A \wedge A \right)  
\end{equation}
where $A$ is a gauge connection for gauge group $SO(N)$ and $k$ is the
coupling constant. The observables in this theory are 
Wilson loop operators: 
\begin{equation}
W_{R_1,R_2, \ldots R_r}[L]~=~ \prod_{i=1}^rTr_{R_i} U [{\cal K}_i]~,
\end{equation}
where $U[{\cal K}_i]=P\left[\exp \oint_{{\cal K}_i} A\right]$ denotes 
the holonomy of the 
$SO(N)$ gauge field $A$ around the component knot ${\cal K}_i$
of a $r$-component link $L$  carrying 
representation $R_i$. The expectation value of these Wilson loop 
operators are the $SO(N)$ link invariants:
\begin{equation}
V_{\Lambda_{R_1},\Lambda_{R_2}, \ldots \Lambda_{R_r}}[L](q,\lambda)= 
\langle W_{R_1,R_2, \ldots R_r}[L] \rangle(q,\lambda)= {\int[{\cal D}A]e^{iS}
 W_{R_1,R_2,\ldots,R_r}[L] \over \int[{\cal D} A]e^{iS}}~, 
 \label {linki} 
\end{equation}
where $\Lambda_{R_i}$'s denote the highest weights of the
representation $R_i$'s. The $SO(N)$ link invariants are 
polynomials in two variables 
\begin{equation}
q=\exp\left({2 \pi i \over k+N-2}\right)~,~ \lambda = q^{N-1}
\end{equation}
involving the coupling constant $k$ and the rank of the gauge group.
These link invariants can be computed using the following two inputs:\\
(i) Any link can be drawn as a closure or plat of braids,~
(ii) The connection between Chern-Simons theory and the Wess-Zumino 
conformal field theory.\\
The invariant for the unknot is equal to the
quantum dimension of the representation $R$  
living on the unknot:
\begin{equation}
V_{\Lambda_R}[U](q,\lambda)=dim_q R ~,\label {unk1}
\end{equation}
where the quantum dimension of the representation $R$ with
highest weight $\Lambda_R$ is given by 
\begin{equation}
dim_q R=\Pi_{\alpha>0}\frac{[\alpha\cdot (\rho+\Lambda_R)]}{[\alpha\cdot\rho]}
\label {unk2}
\end{equation}
where $\alpha$'s are the positive roots and $\rho$ is the Weyl vector
equal to the sum of the fundamental weights of the group $SO(N)$. 
The square bracket refers to the quantum number defined by
\begin{equation}
[x]={\left(q^{x/2}-q^{-x/2} \right) \over \left(q^{1/2}- q^{-1/2} \right)}
\end{equation}

We shall now present the polynomials for various framed knots and links. 
For the unknot $U$ with an arbitrary framing $p$, carrying a representation 
$R$ of $SO(N)$, the polynomial is 
\begin{equation}
V_{\Lambda_R}[0^{(p)}](q,\lambda) =(-1)^{\ell p} q^{(p C_R)}~ V_{\Lambda_R}[U]\label {uinv}
\end{equation}
where $\ell$ refers to the total number of boxes in the Young-Tableau of the
representation $R$ and  
the quadratic Casimir $C_R= \frac{(\Lambda_R + 2 \rho)\cdot \Lambda_R}{2}$ in terms
of Young-Tableau is given by 
\begin{equation}
C_R= {1 \over 2} \left ((N-1) \ell + \ell + \sum_i (l_i^2- 2i l_i) \right)~.
\end{equation} 
Here  $l_i$ denotes the number of boxes in the $i$-th row
of the Young-Tableau of the representation $R$. Eqns.(A.1-A.11) in 
appendix A contain
explicit $p$-framed unknot polynomials for few representations. 

Now, we can write the $SO(N)$ framed knot invariants for torus knots
of the type $(2, 2m+1)$ with framing $[p - (2m+1)]$ as follows:
\begin{equation}
V_{\Lambda_R}[K](q,\lambda)= (-1)^{\ell [p-(2m-1)]} q^{p C_R} \sum_{R_s \in R \otimes R} dim_q R_s 
~(-1)^{\epsilon_s} \left(q^{C_R-C_{R_s}/2}\right)^{2m+1}~,
\end{equation}
where $\epsilon_s= \pm 1$ depending upon whether the representation $R_s$
appears symmetrically or antisymmetrically with respect to the tensor
product $R \otimes R$ in the $SO(N)_k$ Wess-Zumino Witten model. 
Explicit polynomial expression for $p$ framed trefoil for some representations
are presented in appendix A.

Similarly, $SO(N)$ invariants for framed torus links of the type $(2,2m)$
can also be written. For example, the $SO(N)$ invariant for a 
Hopf link with linking number $-1$ and framing numbers 
$p_1$ and $p_2$ on the component knots carrying representations 
$R_1$ and $R_2$ will be 
\begin{equation}
V_{\Lambda_{R_1}, \Lambda_{R_2}}
[H^*(p_1,p_2)](q,\lambda)= (-1)^{\ell_1 p_1 + \ell_2 p_2}q^{p_1 C_{R_1}+p_2 C_{R_2}}
\sum_{R_s \in R_1 \otimes R_2} dim_q R_s q^{C_{R_1} + C_{R_2}- 
C_{R_s}}~.
\end{equation}
We have presented the explicit framed Hopf link polynomials for 
some representations in the appendix A. Using the framed torus 
knot/link invariants, we can write the $SO(N)$ invariants for connected sums.
For example, knot $K=K_1 \# K_2$, where $K_1$ and $K_2$
are framed torus knots, 
\begin{equation}
V_{\Lambda_{R_1}}[K=K_1 \# K_2](q,\lambda)= {1 \over V_{\Lambda_{R_1}}[U](q,\lambda)}
\left(V_{\Lambda_{R_1}}[K_1](q,\lambda)
V_{\Lambda_{R_1}}[K_2](q,\lambda)~\right).
\end{equation}
We can also consider a link $L$ obtained as a connected sum
of a torus knot $K_1$ and a torus link $L_1$. The link
invariant will be
\begin{equation}
V_{\Lambda_{R_1,R_2}}[L=K_1 \# L_1](q,\lambda)= {1 \over V_{\Lambda_{R_1}}[U](q,\lambda)}
\left(V_{\Lambda_{R_1}}[K_1](q,\lambda)V_{\Lambda_{R_1},\Lambda_{R_2}}[L_1](q,\lambda)\right)~.
\end{equation}
In the following section, we will see the
reformulation of $SO(N)$ invariants giving new polynomial invariants.

\section{ Reformulated Link Invariants}

We will briefly review the new polynomial invariants obtained
as a reformulation of link invariants in $U(N)$ Chern-Simons theory.
Then, we address the modified group theoretic equations for the
$SO(N)$ group and show that they also give a similar reformulated
invariants. 

\subsection{$U(N)$ Reformulated Link Invariants}

Ooguri and Vafa showed that the Wilson loop operators in
Chern-Simons theory correspond to certain observables
in the topological string theory giving another piece
of evidence for Gopakumar-Vafa duality
conjecture. The operators in the open topological 
string theory which contains information
about links is given by \cite {ov} 
\begin{equation}
Z(\{U_{\alpha}\}, \{V_{\alpha}\})= \exp\left[ \sum_{\alpha=1}^r \sum_{d=1}
^{\infty} {1 \over d} {\rm Tr} U_{\alpha}^d~ {\rm Tr} V_{\alpha}^d \right]
\label {opr}
\end{equation}
where $U_{\alpha}$ is the holonomy of the gauge connection $A$ around
the component knot ${\cal K}_{\alpha}$ carrying the 
fundamental representation in the $U(N)$ Chern-Simons theory on
$S^3$, and $V_{\alpha}$ is the holonomy of a gauge field $\tilde A$
around the same component knot carrying the fundamental representation
in the $U(M)$ Chern-Simons theory on a Lagrangian three-cycle which 
intersects $S^3$ along the curve ${\cal K}_{\alpha}$. The above
operator can be equivalently represented as
\begin{equation}
Z(\{U_{\alpha}\}, \{V_{\alpha} \})=  1+ \sum_{\{\vec 
k^{(\alpha)}\}} \prod_{\alpha=1}^r 
{1 \over z_{\vec k^{(\alpha)}}} {\bf \gamma}_{\vec k^{(\alpha})} (U_{\alpha})
{\bf \gamma}_{\vec k^{(\alpha)}} (V_{\alpha}) \label {compa}
\end{equation}
where 
\begin{equation}
z_{\vec k^{(\alpha)}}= \prod_j k_j^{(\alpha)}! j^{k_j^{(\alpha)}}~,
{\bf \gamma}_{\vec k^{(\alpha)}} (U_{\alpha}) = \prod_{j=1}^{\infty} 
\left({\rm Tr}U_{\alpha}^j \right)^{k_j^{(\alpha)}}~. \label {defnn}
\end{equation}
Here $\vec k^{(\alpha)}=(k_1^{(\alpha)},k_2^{(\alpha)}, \ldots)$ 
with $\vert \vec k^{(\alpha)} \vert = \sum_j k_j^{(\alpha)}$ and the
sum is over all the vectors $\vec k^{(\alpha)}$ such that
$\sum_{\alpha=1}^r \vert \vec k^{(\alpha)}\vert >0$. 
Using the following group theoretic Frobenius equations,  
\begin{eqnarray}
{\bf \gamma}_{k_1}(U_1) \ldots {\bf \gamma}_{k_r}(U_r)=\sum_{R_1, \ldots R_r} 
\prod_{\alpha=1}^r \chi_{R_{\alpha}}(C(\vec k^{(\alpha)})) 
{\rm Tr}_{R_1}(U_1) \ldots {\rm Tr}_{R_r}(U_r) ~,
\label {frobe}\\
\sum_{\vec k} {1 \over z_{\vec k}} \chi_{R_1}(C(\vec k))) 
\chi_{R_2}(C(\vec k)))= \delta_{R_1 R_2} ~, ~~~~~~~~~~~~~~~~~~~~~~~~~~~~~~~~~~
\label {ortho}
\end{eqnarray}
where $\chi_{R_{\alpha}}(C(\vec k^{(\alpha)}))$'s are characters of the
symmetry group $S_{\ell_{\alpha}}$ with $\ell_{\alpha}=\sum_j j k_j^{(\alpha)}$
and $C(\vec k^{(\alpha)})$ are the conjugacy classes associated
to $\vec k^{(\alpha)}$'s (denoting $k_j^{(\alpha)}$ 
cycles of length $j$), the operator can be shown to be  
\begin{equation}
Z(\{U_{\alpha}\}, \{V_{\alpha} \})=  \sum_{\{R_{\alpha}\}}
\prod_{\alpha=1}^r {\rm Tr}_{R_{\alpha}}(U_{\alpha}) {\rm Tr}_{R_{\alpha}}
(V_{\alpha})~ \label {compb}.
\end{equation}
Ooguri and Vafa have conjectured a specific form for the vacuum expectation
value (vev) of the topological operators (\ref {opr}) for knots\cite {ov} 
invoking the large $N$ topological string duality. This result was further
refined for links\cite{laba} which is generalisable for
framed links \cite {mari} as follows 
\begin{eqnarray} 
\langle Z(\{U_{\alpha}\}, \{V_{\alpha} \}) \rangle_A&=&  
\exp\left[\sum_{d=1}^{\infty}
\sum_{\{R_{\alpha}\}}{1 \over d} 
f_{(R_1,\ldots R_r)}(q^d, \lambda^d)\prod_{\alpha=1}^r {\rm Tr}_{
R_{\alpha}} V_{\alpha}^d \right]~, \label {fexpn}\\
f_{(R_1, R_2, \ldots R_r)} (q, \lambda)&=& \sum_{Q,s} 
{1 \over (q^{1/2}- q^{-1/2})} N_{(R_1, \ldots R_r),Q,s}
q^s \lambda^Q~ \label {fexpo}
\end{eqnarray}
where the suffix $A$ on the vev implies that the expectation
value is obtained by integrating the $U(N)$ gauge fields $A$'s on
$S^3$. Further, for framed links $N_{(R_1, \ldots R_r),Q,s}$ are integers.
In fact, $f_{R_1,R_2, \ldots R_r}(q, \lambda)$ are the $U(N)$
reformulated polynomial invariants involving $U(N)$ Chern-Simons
link invariants.

The general formula for the reformulated polynomial 
invariant $f$ (\ref {fexpo} ) in terms of $U(N)$ framed link 
invariants  $V_{\Lambda_{R_{1j}}, \Lambda_{R_{2j}} \ldots 
\Lambda_{R_{rj}}}^{\{U(N)\}}[L,S^3](q^d, \lambda^d)$ \cite {ramprav}
can be written as \cite{labb}
\begin{eqnarray}
f_{R_1,R_2, \ldots R_r}(q, \lambda)&=& \sum_{d,m=1}^{\infty} (-1)^{m-1}
{\mu(d) \over dm} \sum_{\{\vec k^{(\alpha j)},R_{\alpha j} \}}\times
\nonumber\\
~&~&\prod_{\alpha=1}^r \chi_{R_{\alpha}} \left( C\left( (\sum_{j=1}^m 
\vec k^{(\alpha j)} )_d \right) \right) \prod_{j=1}^m 
{\vert C(\vec k^{(\alpha j)})\vert \over \ell_{\alpha j}!} \times
\nonumber\\
~&~& \chi_{R_{\alpha j}}(C(\vec k^{(\alpha j)})) 
V_{\Lambda_{R_{1j}}, \Lambda_{R_{2j}} \ldots 
\Lambda_{R_{rj}}}^{\{U(N)\}}[L,S^3](q^d, \lambda^d) \label {findd}
\end{eqnarray}
where $\mu(d)$ is the Moebius function defined as follows: 
if $d$ has a prime decomposition ($\{p_i\}$), $d= \prod_{i=1}^a
p_i^{m_i}$, then $\mu(d)=0$ if any of the $m_i$ is greater
than one. If all $m_i=1$, then $\mu(d)=(-1)^a$.
The second sum in the above equation runs over all vectors 
$\vec k^{(\alpha j)}$, with
$\alpha=1 , \ldots r$ and $j=1, \ldots m$, such that 
$\sum_{\alpha=1}^r \vert \vec k^{(\alpha j)} \vert > 0$ for
any $j$ and over representations $R_{\alpha j}$. 
Further $\vec k_d$ is defined as follows: $(\vec k_d)_{di}= k_i$ and has
zero entries for the other components. Therefore, if
$\vec k= (k_1, k_2, \ldots)$, then
\begin{equation}
\vec k_d=(0,\ldots, 0, k_1, 0, \ldots ,0,k_2,0,\ldots),
\end{equation}
where $k_1$ is in the $d$-the entry, $k_2$ in the $2d$-th entry, and so
on. Hence, one can directly evaluate $f$ 
from $U(N)$ framed link invariants and verify the conjecture (\ref{fexpo}).
Further refinement of eqn. (\ref {findd}) revealing the BPS structure
has been presented in \cite{laba}: 
\begin{equation}
f_{R_1,R_2, \ldots R_r}(q, \lambda)= \sum_{R_1', \ldots R_r'}  
M_{R_1, \ldots R_r;R_1',\ldots R_r'} \hat f_{( R_1',\ldots R_r')}(q,\lambda)~.
\end{equation}
where
\begin{eqnarray}
M_{R_1, \ldots R_r;R_1',\ldots R_r'} &=& \sum_{R_1'', \ldots
R_r''} \prod_{\alpha=1}^r C_{R_{\alpha} R_{\alpha}' R_{\alpha}''} 
S_{R_{\alpha}''}(q)~, \label {mrepn}\\
\hat f_{( R_1',\ldots R_r')}(q,\lambda)&=&(q^{-1/2}-q^{1/2})^{r-2}
\sum_{g \geq 0,Q} \hat N_{(R_1', \ldots R_r'),g,Q} (q^{-1/2}-q^{1/2})^{2g}
\lambda^Q~,\label {bps}
\end{eqnarray}
In eqn.(\ref{mrepn}), $R_{\alpha},R_{\alpha}', R_{\alpha}''$ are 
representations of the symmetric group $S_{\ell_{\alpha}}$ which can be
labelled by a Young-Tableau with a total of $\ell_{\alpha}$ boxes 
and $C_{RR'R''}$ are the Clebsch-Gordan coefficients of the symmetric
group. $S_R(q)$ is non-zero only for hook representations. 
For such hook representation having $\ell -d$ boxes in the first row
of the Young Tableau with total $\ell$ boxes, 
$S_R(q)=(-1)^d q^{-(\ell-1)/2+d}$.

\subsection{SO(N) Reformulated Invariants}

As a problem within Chern-Simons field theory, we could 
take the same operator (\ref {opr}) to carry $SO(N)$ holonomy instead
of $U(N)$ holonomy. We shall denote the topological
operator involving $SO(N)$ holonomy as
\begin{equation}
{\tilde Z}(\{\tilde U_{\alpha}\}, \{\tilde V_{\alpha}\})= 
\exp\left[ \sum_{\alpha=1}^r \sum_{d=1}^{\infty} 
{1 \over d} {\rm Tr} \tilde U_{\alpha}^d~ {\rm Tr} \tilde V_{\alpha}^d \right]
\label {opr1}
\end{equation}
where $\tilde U_{\alpha}$ is the holonomy of the gauge connection $A$ around
the component knot ${\cal K}_{\alpha}$ carrying the 
defining representation in the $SO(N)$ Chern-Simons theory on
$S^3$, and $\tilde V_{\alpha}$ is the holonomy of a gauge field $\tilde A$
around the same component knot carrying the defining representation
in the $SO(M)$ Chern-Simons theory on a Lagrangian three-cycle ${\cal C}$ 
which intersects $S^3$ along the curve ${\cal K}_{\alpha}$. In the
context of the duality of the $SO$ Chern-Simons theory to the 
closed string on the orientifold of the resolved conifold \cite {sinha}, 
the gauge group of the Chern-Simons theory on $S^3$ alone has to be
$SO$. Even if  we choose $SO$ gauge group for Chern-Simons theory on 
$S^3$ and $SO$ Chern-Simons theory on Lagrangian cycle ${\cal C}$, 
the results relevant to open topological string amplitude
\cite {vinc2} in the orientifold background will be unaltered. 

Similar to eqn.(\ref {compa}), the above operator can be 
equivalently represented as
\begin{equation}
\tilde Z(\{\tilde U_{\alpha}\}, \{\tilde V_{\alpha} \})=  1+ \sum_{\{\vec 
k^{(\alpha)}\}} \prod_{\alpha=1}^r 
{1 \over z_{\vec k^{(\alpha)}}} {\bf \gamma}_{\vec k^{(\alpha})} 
(\tilde U_{\alpha}) {\bf \gamma}_{\vec k^{(\alpha)}} (\tilde V_{\alpha}) 
\label {compa1} 
\end{equation}
with the usual definitions for $z_{\vec k^{(\alpha)}}$ and 
${\bf \gamma}_{\vec k^{(\alpha)}} (U_{\alpha})$
as given in eqn. (\ref {defnn}).

Now, we need to modify the Frobenius equation which is one of
the main results of the paper. The orthogonality relation
eqn.(\ref {ortho}) remains the same for the $SO$ case.  
The eqn.(\ref {frobe}) will be modified as follows:
\begin{equation}
{\bf \gamma}_{k_1}(U_1) \ldots {\bf \gamma}_{k_r}(U_r)=\sum_{R_1, \ldots R_r} 
\prod_{\alpha=1}^r \chi_{R_{\alpha}}(C(\vec k^{(\alpha)})) 
\hat {{\rm Tr}}_{R_1}(U_1) \ldots \hat {{\rm Tr}}_{R_r}(U_r) ~, \label {frobe1}
\end{equation}
where $\chi_{R_{\alpha}}(C(\vec k^{(\alpha)}))$'s are again characters of the
symmetry group $S_{\ell_{\alpha}}$ with $\ell_{\alpha}=\sum_j j k_j^{(\alpha)}$
and $C(\vec k^{(\alpha)})$ are the conjugacy classes associated
to $\vec k^{(\alpha)}$'s (denoting $k_j^{(\alpha)}$ 
cycles of length $j$). Notice that we have put a `hat' in 
the trace function in the above equation. We shall explain the
meaning of the `hat' by presenting $\hat {{\rm Tr}}_R(U)$ for few 
$SO(N)$ representations (we denote representation $R$ 
by the highest weight $\Lambda_R$ for convenience): 
\begin{eqnarray}
\hat{Tr}_{\lambda^{(1)}}U&=&{Tr}_{\lambda^{(1)}}U \nonumber\\
\hat{Tr}_{2\lambda^{(1)}}U&=&{Tr}_{2\lambda^{(1)}}U+1 \nonumber\\
\hat{Tr}_{\lambda^{(2)}}U&=&{Tr}_{\lambda^{(2)}}U \nonumber\\
\hat{Tr}_{3\lambda^{(1)}}U&=&{Tr}_{3\lambda^{(1)}}U+
{Tr}_{\lambda^{(1)}}U \nonumber\\
\hat{Tr}_{\lambda^{(1)}+\lambda^{(2)}}U&=&{Tr}_{\lambda^{(1)}+\lambda^{(2)}}U+{Tr}_{\lambda^{(1)}}U \nonumber\\
\hat{Tr}_{\lambda^{(3)}}U&=&{Tr}_{\lambda^{(3)}}U\nonumber\\
\hat{Tr}_{4\lambda^{(1)}}U&=&{Tr}_{4\lambda^{(1)}}U+{Tr}_{2\lambda^{(1)}}U+1 \nonumber\\
\hat{Tr}_{2\lambda^{(1)}+\lambda^{(2)}}U&=&{Tr}_{2\lambda^{(1)}+\lambda^{(2)}}U+{Tr}_{2\lambda^{(1)}}U+{Tr}_{\lambda^{(2)}}U\nonumber\\
\hat{Tr}_{2\lambda^{(2)}}U&=&{Tr}_{2\lambda^{(2)}}U+{Tr}_{2\lambda^{(1)}}U+1\nonumber\\
\hat{Tr}_{\lambda^{(1)}+\lambda^{(3)}}U&=&{Tr}_{\lambda^{(1)}+\lambda^{(3)}}U+{Tr}_{\lambda^{(2)}}U\nonumber\\
\hat{Tr}_{\lambda^{(4)}}U&=&{Tr}_{\lambda^{(4)}}U~.\label {hattr}
\end{eqnarray}
In principle, $\hat {Tr}$ can be derived for arbitrary $SO(N)$ representation
with highest weight $\Lambda_R= \sum_{i=1}n^{(R)}_i \lambda^{(i)}.$ 
In the next section, we will use the above set of
$\hat {Tr}$ (\ref {hattr}) for obtaining explicit results
 on topological open-string amplitudes. 

Using the eqns.(\ref {frobe1}, \ref {ortho}), it is not difficult to see that
the $S0$ topological operator (\ref {compa1}) is equivalent to
\begin{equation}
\tilde Z(\{\tilde U_{\alpha}\}, \{\tilde V_{\alpha} \})=   
\sum_{\{R_{\alpha}\}}
\prod_{\alpha=1}^r \hat {\rm Tr}_{R_{\alpha}}(\tilde U_{\alpha}) 
\hat {\rm Tr}_{R_{\alpha}}
(\tilde V_{\alpha})~ \label {compb1}.
\end{equation}
Similar to Ooguri-Vafa conjecture, we propose the following
conjecture for the operator (\ref {opr1}) involving $SO$ holonomy:\\
\underline{\bf Conjecture 1}:
\begin{eqnarray} 
e^{{\cal F}(\{{\tilde V}_{\alpha}\})}= \langle \tilde Z(\{\tilde U_{\alpha}\}, 
\{\tilde V_{\alpha} \})\rangle_A&=&   
\exp\left[\sum_{d=1}^{\infty} \sum_{\{R_{\alpha}\}}{1 \over d} 
g_{(R_1,\ldots R_r)}(q^d, \lambda^d)\prod_{\alpha=1}^r {\rm Tr}_{
R_{\alpha}} \tilde V_{\alpha}^d \right]~, \label {fexpn1}\\
g_{(R_1, R_2, \ldots R_r)} (q, \lambda)&=& \sum_{Q,s} 
{1 \over (q^{1/2}- q^{-1/2})} \tilde N_{(R_1, \ldots R_r),Q,s}
q^s \lambda^Q~ \label {fexpo1}
\end{eqnarray}
where the suffix $A$ on the vev implies that the expectation
value is obtained by integrating the $SO(N)$ gauge fields $A$'s on
$S^3$ and $\tilde N_{(R_1, \ldots R_r),Q,s}$ in eqn.(\ref {fexpo1}) are 
integers. We have introduced $\cal F$$(\{{\tilde V}_{\alpha} \})$
which we call as open-string partition function. Incidentally,
${\cal F}$$(\{{\tilde V}_{\alpha} \})$ is a sum of oriented string
partition function(untwisted sector) and unoriented string partition function
(twisted sector) as presented in ref.\cite{vinc2}.

The function $g_{R_1, R_2 \ldots R_r}(q, \lambda)$ are the 
$SO(N)$ reformulated polynomial invariants involving framed link invariants
in $SO(N)$ Chern-Simons theory.
It is easy to see that the eqn.(\ref {findd}) can be accordingly 
modified for $SO$ group involving the expectation
value of $\hat {Tr}_R\tilde U$ (\ref {hattr}) as follows:
\begin{eqnarray}
g_{R_1,R_2, \ldots R_r}(q, \lambda)&=& \sum_{d,m=1}^{\infty} (-1)^{m-1}
{\mu(d) \over dm} \sum_{\{\vec k^{(\alpha j)},R_{\alpha j} \}}\times
\nonumber\\
~&~&\prod_{\alpha=1}^r \chi_{R_{\alpha}} \left( C\left( (\sum_{j=1}^m 
\vec k^{(\alpha j)} )_d \right) \right) \prod_{j=1}^m 
{\vert C(\vec k^{(\alpha j)})\vert \over \ell_{\alpha j}!} \times
\nonumber\\
~&~& \chi_{R_{\alpha j}}(C(\vec k^{(\alpha j)})) \langle 
\prod_{\alpha=1}^r \hat {Tr}_{R_{\alpha j}}\tilde U_{\alpha}[{\cal K}_{\alpha}]
\rangle (q^d, \lambda^d) \label {findde} 
\end{eqnarray}
where the definitions of $\mu(d)$, $\vec k^{(\alpha j)}$ and $\vec k_d$ 
are same as defined in the previous subsection. 
Substituting the $\hat {Tr}$ (\ref {hattr}) and rewriting
in terms of $SO(N)$ framed link invariants (\ref {linki}), 
we have explicitly verified that the $SO(N)$ reformulated
invariant obeys the conjectured eqn.(\ref {fexpo1}) for
many framed knots and framed links. This is one of the
non-trivial results of the paper which we present in the next section
and in appendix B.

Using eqn. (\ref {mrepn}), we can rewrite
the $SO(N)$ reformulated polynomial invariants 
$g_{R_1,R_2, \ldots R_r}(q, \lambda)$ as 
\begin{equation}
g_{R_1,R_2, \ldots R_r}(q, \lambda)= \sum_{R_1', \ldots R_r'}  
M_{R_1, \ldots R_r;R_1',\ldots R_r'} \hat g_{( R_1',\ldots R_r')}(q,\lambda)~.
\end{equation}
Unfortunately, $\hat g_{R_1, \ldots R_r}(q,\lambda)$ does not have
a BPS structure like the one given in eqn.(\ref {bps}) 
for $U(N)$ holonomy. This has also been extensively studied 
in the works of Bouchard et al \cite {vinc2} where they 
conjecture an equation for $\hat g_R(q, \lambda)$ corresponding
to knots in standard framing as follows:
\begin{equation} 
\frac{1}{2}\left(\hat{g}_{R}(q,{\lambda}^{\frac{1}{2}})-
(-1)^{\ell(R)}\hat{g}_{R}(q,-{\lambda}^{\frac{1}{2}})\right)=
\sum_{g,\beta} N_{R,g,\beta}^{c=1} 
{\left(q^{\frac{1}{2}}-q^{-\frac{1}{2}}\right)}^{2g}{\lambda}^{\beta}~.\label
{moda}
\end{equation} 
In this equation, $N_{R,g,\beta}^{c=1}$ are 
BPS invariants corresponding to unoriented open string amplitudes with one 
cross-cap. The above conjecture can be generalised for arbitrary 
$r$-component framed links $[L, {\bf p}]$ where 
${\bf p}=(p_1, p_2 \ldots p_r)$ denotes the framing
numbers $p_i$'s on the component knots ${\cal K}_i$'s. For such 
$r$-component framed links $[L, {\bf p}]$, 
we propose the following conjecture:\\
\underline{\bf Conjecture 2} 
\begin{eqnarray} 
\frac{1}{2}\left(\hat{g}_{R_1,R_2 \ldots R_r}(q,{\lambda}^{\frac{1}{2}})-
(-1)^{\sum_{\alpha=1}^r\ell(R_{\alpha})(p_{\alpha}+1)}
\hat{g}_{R_1,R_2, \ldots R_r}(q,-{\lambda}^{\frac{1}{2}})\right)=\nonumber\\
~~~~~~~~~~~~~~~~~~~~~\sum_{g,\beta} N_{R_1,R_2, \ldots R_r,g,\beta}^{c=1} 
{\left(q^{\frac{1}{2}}-q^{-\frac{1}{2}}\right)}^{2g+r-1}{\lambda}^{\beta} \label{mod1}
\end{eqnarray}
We have verified the above conjecture for many framed knots and framed 
two component links. In section 5, we have presented 
$N_{R_1,R_2, \ldots R_r,g,\beta}^{c=1}$ for some framed knots
and framed two component links.

\section{Explicit Computation of $SO(N)$ reformulated invariants 
$g_{R_1,R_2, \ldots R_r}(q,\lambda)$}

In this section we compute the functions $g_{R_1, \ldots
R_r}(q,\lambda)$ for various nontrivial framed knots and links and
show that they obey the conjectured  
form (\ref {fexpo1}). We shall denote the representations $R_i$'s in 
$g_{R_1, \ldots R_r}(q, \lambda)$ by their highest weights $\Lambda_{R_i}$'s. 

\begin{itemize}
\item For unknot in standard framing, the reformulated invariant is
{\bf non-zero only for the defining representation}: 
\begin{equation} 
V_{\lambda^{(1)}}[U](q,\lambda)=g_{\lambda^{(1)}}(q,\lambda)=
\frac{1}{q - 1}\left[q - 1+q^{1/2}\,
\left( -1 + \lambda  \right){\lambda }^{-1/2}\right]~.
\end{equation} 
This simplifies the form for open-string
partition function ${\cal F}$$(\{{\tilde V}_{\alpha}\})$ 
in eqn. (\ref {fexpn1}) as follows:
\begin{equation}
{\cal F}(\tilde V) = \sum_d {1 \over d} \left( 1 + {\lambda^{d/2}
- \lambda^{-d/2}  \over q^{d/2} - q^{-1/2}}\right) {\rm Tr} \tilde V^d
\end{equation}
\item For unknot with arbitrary framing $p$
\begin{eqnarray} 
g_{\lambda^{(1)}}(q,\lambda)&=&{\left( -1 \right) }^p\,
{\lambda }^{p/2}
  \left( 1 + \frac{{q^{1/2}}\,
       \left( -1 + \lambda  \right) }{\left( -1 + 
         q \right) \,{{\lambda }^{1/2}}} \right) \\ \nonumber\\
g_{2\lambda^{(1)}}(q,\lambda)&=& \frac{1}{2\,{\left( -1 + q \right) }^2\,
     \left( 1 + q \right) }\left[2\,{\left( -1 + q \right) }^2\,
     \left( 1 + q \right)\right.\nonumber\\
&&+{\lambda }^{-1 + p}\,
     \left( 2\,q^p \left( -1 + \lambda  \right) \,
        \left( -q + {q^{1/2}}
           \left( -1 + q^2 \right) \,
           {\lambda }^{1/2} + q^2\,\lambda  \right)\right.\nonumber\\
      && + {\left( -1 \right) }^p
        \left( -\left( {\left( -1 \right) }^p\,
             \left( 1 + q \right) 
             {\left( {q^{1/2}} + 
                 {\lambda }^{1/2} \right) }^2
             {\left( -1 + 
                 {q^{1/2}}\,{\lambda }^{1/2}
                 \right) }^2 \right) \right.\nonumber\\ 
      &&\left.\left.\left. - \left( -1 + q \right) \,
           \left( q + \lambda  \right) \,
           \left( -1 + q\,\lambda  \right)  \right) 
       \right) \right]\\ \nonumber\\
g_{\lambda^{(2)}}(q,\lambda)&=&\frac{1}{2\,{\left( -1 + q \right) }^2\,
    \left( 1 + q \right) }\left[{\lambda }^{-1 + p}\,
    \left( 2\,
       \left( q^{3/2} + 
         {\lambda }^{1/2} \right) \left( -1 +
         {q}^{1/2}\,{\lambda }^{1/2} \right)\right.\right.\nonumber \\
&&    q^{{1/2} - p}\,  \left( -1 + \lambda  \right)  - 
            \left( 1 + q \right) 
      {\left( {q}^{1/2} + 
                {\lambda }^{1/2} \right) }^2\,
            {\left( -1 + 
                {q}^{1/2}\,{\lambda }^{1/2}
                \right) }^2   \nonumber \\
&&\left.\left.+ (-1)^p
         \left( -1 + q \right) \,
          \left( q + \lambda  \right) \,
          \left( -1 + q\,\lambda  \right) 
      \right) \right]  
\end{eqnarray}
\begin{eqnarray}
g_{3\lambda^{(1)}}(q,\lambda)&=&\frac{1}{{\left( -1 + q \right) }^3
    \left( 1 + q \right) 
    \left( 1 + q + q^2 \right) }\left[{\left( -1 \right) }^p
    \left( -1 + q^p \right) 
    \left( q^{1/2} + {\lambda }^{1/2} \right)\right. \nonumber \\ 
&&    \left( -1 + q^{1/2}{\lambda }^{1/2}
      \right) \left( -1 + \lambda  \right) q^{1/2}
    {\lambda }^
     {{3\left( -1 + p \right) }/2}
    \left( {\lambda }^{1/2} -
      q^{3/2}
       \left( -1 + \right.\right.\nonumber \\
&& \left.q
          \left( -1 + \lambda  \right)  + q^{5/2}{\lambda }^{1/2} +
         \lambda  \right)  +
      q^p\left( -q^{1/2} +
         q\left( -1 + q^2 \right) 
          {\lambda }^{1/2} + \right.\nonumber \\
&&\left.\left.\left.  q^{7/2}\,\lambda  \right)   + q^{2\,p}\,\left( -q^{1/2} +
         q\,\left( -1 + q^2 \right) \,
          {\lambda }^{1/2} +
         q^{7/2}\,\lambda  \right)  \right)\right]
\end{eqnarray}
\begin{eqnarray}
g_{\lambda^{(1)}+\lambda^{(2)}}(q,\lambda)&=& \frac{-1}{{\left( -1 +
          q \right) }^3\,\left( 1 + q \right) }\left[{\left( -1 \right) }^p\,
      q^{1/2 - p}\,
      \left( -1 + q^p \right) \,
      \left( q^{1/2} + {\lambda }^{1/2} \right)\right.\nonumber \\
&&   \left( -1 +
        q^{1/2}\,{\lambda }^{1/2} \right) \,
      \left( -1 + \lambda  \right) \,
      {\lambda }^
       {{3\,\left( -1 + p \right) }/2}\,
      \left( 
        q^p\,\left( q^{1/2} +
           {\lambda }^{1/2} \right) \right.\nonumber \\
&&\left.\left.         \left( -1 +
           q^{3/2}\,{\lambda }^{1/2}
           \right) + q^{3/2} + {\lambda }^{1/2} -
        q^2\,{\lambda }^{1/2} -
        q^{1/2}\,\lambda  \right) \right]\\ \nonumber \\
g_{\lambda^{(3)}}(q,\lambda)&=&\frac{1}{{\left( -1 +
        q \right) }^3\,\left( 1 + q \right) \,
    \left( 1 + q + q^2 \right) }\left[{\left( -1 \right) }^p\,
    q^{1/2 - 3\,p}\,
    \left( -1 + q^p \right)  \right.\nonumber \\
&&  \left( q^{1/2} + {\lambda }^{1/2} \right) \left( -1 + q^{1/2}\,{\lambda }^{1/2}
      \right) \,\left( -1 + \lambda  \right) \,
    {\lambda }^
     {{3\,\left( -1 + p \right) }/2} \nonumber \\
&&    \left( q^{7/2} -
      q\left( -1 + q^2 \right) 
       {\lambda }^{1/2} - q^{1/2}\,\lambda + q^p\,\left( q^{7/2} - \right.\right.\nonumber \\
&& \left.
         q\,\left( -1 + q^2 \right) \,
          {\lambda }^{1/2} - q^{1/2}\,\lambda
         \right)  + q^{2\,p}\,
       \left( -\left( q^{3/2}\,
            \left( 1 + q \right)  \right) \right.\nonumber \\ 
&&\left.\left.\left.       +  \left( -1 + q^4 \right) \,
          {\lambda }^{1/2} +
         q^{3/2}\,\left( 1 + q \right) \,
          \lambda  \right)  \right) \right]
\end{eqnarray} 
Substituting values for $p$, the above equations reduce to the conjectured
form (\ref {fexpo1}).
\begin{enumerate}
\item For unknot with framing $p=1$, we get
\begin{eqnarray} 
g_{\lambda^{(1)}}(q,\lambda)&=&\frac{-1}{q - 1}\left[(q - 1)\lambda^{1/2}
+q^{1/2}\, \left( -1 + \lambda  \right)\right]\nonumber\\
g_{2\lambda^{(1)}}(q,\lambda)&=&\frac{1}{q - 1}
	\left[\left( -1 + q^{1/2}\,{\lambda }^{1/2} \right) \,
    \left( 1 + q\,\left( -1 + \lambda  \right)  +
      q^{1/2}\,{\lambda }^{3/2} \right) \right]\nonumber\\
g_{\lambda^{(2)}}(q,\lambda)&=& \frac{-1}{q - 1}\left[\left( 
       - q^{-1/2}{\lambda }^{1/2} + \lambda \right) 
      \left( -1 + q + q^{1/2}{\lambda }^{1/2} +
        \lambda  \right) \right] \nonumber
\end{eqnarray}  
\item For unknot with framing two 
\begin{eqnarray} 
g_{\lambda^{(1)}}(q,\lambda)&=&\frac{1}{q - 1}\left[(q - 1) \lambda + 
q^{1/2}\, \left( -1 + \lambda  \right)\,{\lambda }^{1/2}\right]\nonumber\\
g_{2\lambda^{(1)}}(q,\lambda)&=&\frac{1}{q-1}\left[\left( -1 + \lambda  
\right) \,
    \{ 1 - q + \lambda  - 2\,q\,\lambda  +
      q^{1/2} \times \, \right .\nonumber \\
~&~&\left . \left( -1 + q^2 \right) \,
       {\lambda }^{3/2} + q^2\,{\lambda }^2
      \} \right] \nonumber\\
g_{\lambda^{(2)}}(q,\lambda)&=&\frac{-1}{q-1}\left[q^{-3/2}\left( q^{3/2} +
        {\lambda }^{1/2} \right) \,
      \left( -1 + q^{1/2}\,{\lambda }^{1/2} \right) \,
      \left( -1 + \lambda  \right) \,\lambda \right]
\end{eqnarray} 
\end{enumerate}
\item We have presented the reformulated invariants for 
few framed knots and two component links in appendix B.
\end{itemize}

\section{$N_{(R_1, \ldots R_r),g,Q}^{c=1}$ Computation}

We shall now  compute the integer coefficients (\ref {mod1}) corresponding
to cross-cap $c=1$ unoriented open string amplitude obtained
from $SO(N)$ reformulated invariants for various framed knots and framed
links.
\subsection{Framed Knots}
\begin{enumerate}
\item For unknot with zero framing,
the only non zero coefficient is $N_{\lambda^{(1)},0,0}^{c=1}=1$. 
\item  For unknot with framing $p=1$
\begin{equation} 
N_{\lambda^{(1)},0,1/2}^{c=1}=-1
\end{equation} 

\begin{center} 
\begin{tabular}{|r|cc|}\hline
 & $\beta$=1/2 & 3/2 \\ \hline
g=0 & 1 & -1 \\ \hline
\end{tabular} 
\hspace{1in}
\begin{tabular}{|r|cc|}\hline
 & $\beta$=3/2 & 5/2 \\ \hline
g=0 & 1 & -1 \\ \hline
\end{tabular} 

$N_{\lambda^{(2)},g,\beta}^{c=1}$\hspace{1.75in}
$N_{\lambda^{(1)}+\lambda^{(2)},g,\beta}^{c=1}$
\end{center}  

\begin{center}
\begin{tabular}{|r|ccc|}  \hline
\em  & $\beta$=1/2 & 3/2 & 5/2\\ \hline
g=0 & -1 & 4 & -3\\
1 & 0 & 1 & -1\\ \hline
\end{tabular}

$N_{\lambda^{(3)},g,\beta}^{c=1}$
\end{center}

\item For unknot with framing $p=2$:
\begin{equation}
N_{\lambda^{(1)},0,1}^{c=1}=1
\end{equation}

\begin{center} 
\begin{tabular}{|r|cc|}\hline
 & $\beta$=3/2 & 5/2 \\ \hline
g=0 & 1 & -1 \\ \hline
\end{tabular}

$N_{2\lambda^{(1)},0,3/2}^{c=1}$
\end{center}  

\begin{center}
\begin{tabular}{|r|cc|}  \hline
\em  & $\beta$=3/2 & 5/2 \\ \hline
g=0 & 3 & -3\\
1 & 1 & -1\\ \hline
\end{tabular}
\hspace{1in}
\begin{tabular}{|r|ccc|}  \hline
\em  & $\beta$=2 & 3 & 4\\ \hline
g=0 & 1 & -4 & 3\\
1 & 0 & -1 & 1\\ \hline
\end{tabular}

$N_{\lambda^{(2)},g,\beta}^{c=1}$\hspace{2in}
$N_{3\lambda^{(1)},g,\beta}^{c=1}$
\end{center}

\begin{center}
\begin{tabular}{|r|ccc|}  \hline
\em  & $\beta$=2 & 3 & 4\\ \hline
g=0 & 9 & -28 & 19\\
1 & 6 & -27 & 21\\
2 & 1 & -9 & 8 \\
3 & 0 & -1 & 1 \\ \hline
\end{tabular}
\hspace{1in}
\begin{tabular}{|r|ccc|}  \hline
\em  & $\beta$=2 & 3 & 4\\ \hline
g=0 & 13 & -36 & 23\\
1 & 16 & -57 & 41\\
2 & 7 & -36 & 29 \\
3 & 1 & -10 & 9 \\
4 & 0 & -1 & 1\\ \hline
\end{tabular}
                                                                                
$N_{\lambda^{(1)}+\lambda^{(2)},g,\beta}^{c=1}$\hspace{2in}
$N_{\lambda^{(3)},g,\beta}^{c=1}$
\end{center}

\item For trefoil knot in standard framing, the results
are agreeing with the tables given in Ref. \cite {vinc2}.

\item For trefoil knot with framing $p=1$

\begin{center}
\begin{tabular}{|r|ccc|}  \hline
\em  & $\beta$=3/2 & 5/2 & 7/2\\ \hline
g=0 & -3 & 3 & -1\\
1 & -1 & 1 & 0\\ \hline
\end{tabular}

$N_{\lambda^{(1)},g,\beta}^{c=1}$
\end{center}

\begin{center}
\begin{tabular}{|r|ccccc|}  \hline
\em  & $\beta$=5/2 & 7/2 & 9/2 & 11/2 & 13/2\\ \hline
g=0 & 16 & -69 & 111 & -79 & 21\\
1 & 20 & -146 & 307 & -251 & 70\\
2 & 8 & -128 & 366 & -330 & 84 \\
3 & 1 & -56 & 230 & -220 & 45\\
4 & 0 & -12 & 79 & -78 & 11\\ 
5 & 0 & -1 & 14 & -14 & 1\\
6 & 0 & 0 & 1& -1 & 0 \\\hline
\end{tabular}

$N_{2\lambda^{(1)},g,\beta}^{c=1}$
\end{center}

\begin{center}
\begin{tabular}{|r|ccccc|}  \hline
\em  & $\beta$=5/2 & 7/2 & 9/2 & 11/2 & 13/2\\ \hline
g=0 & 30 & -114 & 167 & -111 & 28\\
1 & 55 & -311 & 587 & -457 & 126\\
2 & 36 & -367 & 912 & -791 & 210 \\
3 & 10 & -230 & 770 & -715 & 165\\
4 & 1 & -79 & 376 & -364 & 66\\
5 & 0 & -14 & 106 & -105 & 13\\
6 & 0 & -1 & 16 & -16 & 1 \\
7 & 0 & 0 & 1 & -1 & 0\\\hline
\end{tabular}
                                                                                
$N_{\lambda^{(2)},g,\beta}^{c=1}$
\end{center}

\item For trefoil knot with framing $p=2$

\begin{center}
\begin{tabular}{|r|ccc|}  \hline
\em  & $\beta$=2 & 3 & 4\\ \hline
g=0 & 3 & -3 & 1\\
1 & 1 & -1 & 0\\ \hline
\end{tabular}
                                                                                
$N_{\lambda^{(1)},g,\beta}^{c=1}$
\end{center}

\begin{center}
\begin{tabular}{|r|rrrrr|}  \hline
\em  & $\beta$=7/2 & 9/2 & 11/2 & 13/2 & 15/2\\ \hline
g=0 & 30 & -114 & 167 & -111 & 28\\
1 & 55 & -311 & 587 & -457 & 126\\
2 & 36 & -367 & 912 & -791 & 210 \\
3 & 10 & -230 & 770 & -715 & 165\\
4 & 1 & -79 & 376 & -364 & 66\\
5 & 0 & -14 & 106 & -105 & 13\\
6 & 0 & -1 & 16 & -16 & 1 \\
7 & 0 & 0 & 1 & -1 & 0\\\hline
\end{tabular}
                                                                                
$N_{2\lambda^{(1)},g,\beta}^{c=1}$
\end{center}

\begin{center}
\begin{tabular}{|r|rrrrr|}  \hline
\em  & $\beta$=7/2 & 9/2 & 11/2 & 13/2 & 15/2\\ \hline
g=0 & 50 & -174 & 237 & -149 & 36\\
1 & 125 & -601 & 1042 & -776 & 210\\
2 & 120 & -919 & 2046 & -1709 & 462 \\
3 & 55 & -771 & 2222 & -2001 & 495\\
4 & 12 & -376 & 1443 & -1365 & 286\\
5 & 1 & -106 & 574 & -560 & 91\\
6 & 0 & -16 & 137 & -136 & 15 \\
7 & 0 & -1 & 18 & -18 & 1\\
8 & 0 & 0 & 1 & -1 & 0\\\hline
\end{tabular}
                                                                                
$N_{\lambda^{(2)},g,\beta}^{c=1}$
\end{center}

\item For connected sum Trefoil \# Trefoil with zero framing

\begin{center} 
\begin{tabular}{|r|cccc|}\hline
\em & $\beta$=2 & 3 & 4 & 5 \\ \hline
g=0 & 8 & -14 & 9 & -2 \\
1 & 6 & -11 & 6 & -1 \\
2 & 1 & -2 & 1 & 0 \\ \hline
\end{tabular} 

$N_{\lambda^{(1)},g,\beta}^{c=1}$
\end{center}  

\begin{center} 
\begin{tabular}{|r|rrrrrrrr|} \hline
\em & $\beta$=7/2 & 9/2 & 11/2 & 13/2 & 15/2 & 17/2 & 19/2 & 21/2 \\ \hline
g=0 & 143 & -831 & 1950 & -2366 & 1561 & -525 & 66 & 2\\
1 & 404 & -3144 & 8854 & -11819 & 7544 & -1488 & -596 & 245 \\
2 & 464 & -5419 & 19211 & -28097 & 14046 & 6348 & -9194 & 2641 \\
3 & 277 & -5379 & 25184 & -40255 & 6296 & 44160 & -41756 & 11473 \\
4 & 90 & -3292 & 21666 & -38551 & -18588 & 110890 & -98450 & 26235 \\
5 & 15 & -1256 & 12654 & -26241 & -38613 & 159091 & -141400 & 35750 \\
6 & 1 & -290 & 5048 & -13093 & -36589 & 147270 & -133378 & 31031 \\
7 & 0 & -37 & 1352 & -4787 & -21053 & 92681 & -85919 & 17763 \\
8 & 0 & -2 & 232 & -1243 & -7860 & 40544 & -38455 & 6784 \\
9 & 0 & 0 & 23 & -215 & -1917 & 12353 & -11954 & 1710 \\
10 & 0 & 0 & 1 & -22 & -295 & 2574 & -2531 & 273 \\
11 & 0 & 0 & 0 & -1 & -26 & 350 & -348 & 25 \\
12 & 0 & 0 & 0 & 0 & -1 & 28 & -28 & 1 \\
13 & 0 & 0 & 0 & 0 & 0 & 1 & -1 & 0 \\ \hline
\end{tabular} 

$N_{2\lambda^{(1)},g,\beta}^{c=1}$
\end{center}  

\begin{center} 
\begin{tabular}{|r|rrrrrrrr|} \hline
\em & $\beta$=7/2 & 9/2 & 11/2 & 13/2 & 15/2 & 17/2 & 19/2 & 21/2 \\ \hline
g=0 & 227 & -1237 & 2756 & -3206 & 2045 & -671 & 84 & 2 \\
1 & 801 & -5621 & 14872 & -19187 & 12269 & -2861 & -564 & 291 \\
2 & 1190 & -11771 & 38341 & -54346 & 29773 & 5595 & -12485 & 3697 \\
3 & 955 & -14403 & 59796 & -92245 & 27489 & 67819 & -68353 & 18942 \\
4 & 444 & -11132 & 61614 & -104212 & -18925 & 211571 & -190697 & 51337 \\
5 & 119 & -5578 & 43750 & -83517 & -79482 & 364896 & -323843 & 83655 \\
6 & 17 & -1803 & 21761 & -49317 & -100043 & 404876 & -363462 & 87971 \\
7 & 1 & -362 & 7561 & -21758 & -73516 & 307780 & -281842 & 62136 \\
8 & 0 & -41 & 1795 & -7097 & -35330 & 165164 & -154547 & 30056 \\
9 & 0 & -2 & 277 & -1653 & -11446 & 63250 & -60402 & 9976 \\
10 & 0 & 0 & 25 & -258 & -2483 & 17202 & -16719 & 2233 \\
11 & 0 & 0 & 1 & -24 & -346 & 3248 & -3201 & 322 \\
12 & 0 & 0 & 0 & -1 & -28 & 405 & -403 & 27 \\
13 & 0 & 0 & 0 & 0 & -1 & 30 & -30 & 1 \\
14 & 0 & 0 & 0 & 0 & 0 & 1 & -1 & 0 \\ \hline
\end{tabular} 

$N_{\lambda^{(2)},g,\beta}^{c=1}$
\end{center}  
\end{enumerate}

\subsection{Framed Links}
\begin{enumerate}
\item {Hopf Link}

We take Hopf Link $H(p_1,p_2)$ with linking number -1 and framing on the two
component knots as $p_1$ and $p_2$. The integers $N_{(R_1,R_2),g,\beta}^{c=1}$
for various combinations of $p_1$ and $p_2$ are tabulated below.

$\underline{p_1=0=p_2}$
\begin{equation} 
N_{(\lambda^{(1)},\,\lambda^{(1)}),0,1/2}^{c=1}=1
\end{equation} 

\begin{center} 
\begin{tabular}{|r|cc|} \hline
\em & $\beta$=-1 & 0\\ \hline
g=0 & -1 & 1\\ \hline
\end{tabular} 
\hspace{1in}
\begin{tabular}{|r|cc|} \hline
\em & $\beta$=0 & 1\\ \hline
g=0 & 1 & -1\\ \hline
\end{tabular} 

$N_{(2\lambda^{(1)},\,\lambda^{(1)}),g,\beta}^{c=1}$\hspace{1.5in}
$N_{(\lambda^{(2)},\,\lambda^{(1)}),g,\beta}^{c=1}$  
\end{center}  

\begin{center} 
\begin{tabular}{|r|cc|} \hline
\em & $\beta$=-3/2 & -1/2\\ \hline
g=0 & -1 & 1\\ \hline
\end{tabular}
\hspace{0.25in}
\begin{tabular}{|r|cc|} \hline
\em & $\beta$=-1/2 & 1/2\\ \hline
g=0 & 1 & -1\\ \hline
\end{tabular}
\hspace{0.25in}
\begin{tabular}{|r|cc|} \hline
\em & $\beta$=1/2 & 3/2\\ \hline
g=0 & -1 & 1\\ \hline
\end{tabular} 

$N_{(3\lambda^{(1)},\,\lambda^{(1)}),g,\beta}^{c=1}$\hspace{0.9in}
$N_{(\lambda^{(1)}+\lambda^{(2)},\,\lambda^{(1)}),g,\beta}^{c=1}$\hspace{0.9in}
$N_{(\lambda^{(3)},\,\lambda^{(1)}),g,\beta}^{c=1}$
\end{center}  

$\underline{p_1=1=p_2}$

\begin{center} 
\begin{tabular}{|r|cc|}\hline
\em & $\beta$=1/2 & 3/2\\ \hline
g=0 & -1 & 1\\ \hline
\end{tabular}
\hspace{0.25in}
\begin{tabular}{|r|cc|}\hline
\em & $\beta$=3/2 & 5/2\\ \hline
g=0 & -1 & 1\\ \hline
\end{tabular}
\hspace{0.25in}
\begin{tabular}{|r|ccc|}\hline
\em & $\beta$=1/2 & 3/2 & 5/2\\ \hline
g=0 & 1 & -5 & 4\\ 
1 & 0 & -1 & 1\\ \hline
\end{tabular}

$N_{(\lambda^{(1)},\,\lambda^{(1)}),g,\beta}^{c=1}$\hspace{0.85in}
$N_{(2\lambda^{(1)},\,\lambda^{(1)}),g,\beta}^{c=1}$\hspace{0.85in}
$N_{(\lambda^{(2)},\,\lambda^{(1)}),g,\beta}^{c=1}$
\end{center}

\begin{center} 
\begin{tabular}{|r|cc|}\hline
\em & $\beta$=5/2 & 7/2\\ \hline
g=0 & -1 & 1\\ \hline
\end{tabular} 

$N_{(3\lambda^{(1)},\,\lambda^{(1)}),g,\beta}^{c=1}$
\end{center}  

\begin{center} 
\begin{tabular}{|r|ccc|}\hline
\em & $\beta$=3/2 & 5/2 & 7/2\\ \hline
g=0 & 6 & -19 & 13\\
1 & 1 & -8 & 7\\
2 & 0 & -1 & 1\\ \hline
\end{tabular} 
\hspace{1in}
\begin{tabular}{|r|cccc|}\hline
\em & $\beta$=1/2 & 3/2 & 5/2 & 7/2 \\ \hline
g=0 & -1 & 15 & -36 & 22\\
1 & 0 & 7 & -29 & 22\\
2 & 0 & 1 & -9 & 8\\
3 & 0 & 0 & -1 & 1\\ \hline
\end{tabular}

$N_{(\lambda^{(1)}+\lambda^{(2)},\,\lambda^{(1)}),g,\beta}^{c=1}$\hspace{2in}
$N_{(\lambda^{(3)},\,\lambda^{(1)}),g,\beta}^{c=1}$
\end{center}  

$\underline{p_1=2=p_2}$

\begin{center}
\begin{tabular}{|r|cc|}\hline
\em & $\beta$ = 3/2 & 5/2\\ \hline
g=0 & -1 & 1\\ \hline
\end{tabular}
\hspace{0.5in}
\begin{tabular}{|r|ccc|}\hline
\em & $\beta$=2 & 3 & 4 \\ \hline
g=0 & -1 & 5 & -4\\
1 & 0 & 1 & -1\\ \hline
\end{tabular}

$N_{(\lambda^{(1)},\,\lambda^{(1)}),g,\beta}^{c=1}$\hspace{1in}
$N_{(2\lambda^{(1)},\,\lambda^{(1)}),g,\beta}^{c=1}$
\end{center}

\begin{center} 
\begin{tabular}{|r|ccc|}\hline
\em & $\beta$=2 & 3 & 4 \\ \hline
g=0 & -4 & 13 & -9\\
1 & -1 & 7 & -6\\ 
2 & 0 & 1 & -1 \\\hline
\end{tabular}

$N_{(\lambda^{(2)},\,\lambda^{(1)}),g,\beta}^{c=1}$
\end{center}  

\begin{center} 
\begin{tabular}{|r|cccc|}\hline
\em & $\beta$=5/2 & 7/2 & 9/2 & 11/2\\ \hline
g=0 & -1 & 15 & -36 & 22\\
1 & 0 & 7 & -29 & 22\\
2 & 0 & 1 & -9 & 8\\
3 & 0 & 0 & -1 & 1\\ \hline
\end{tabular}

$N_{(3\lambda^{(1)},\,\lambda^{(1)}),g,\beta}^{c=1}$
\end{center}  

\begin{center} 
\begin{tabular}{|r|cccc|}\hline
\em & $\beta$=5/2 & 7/2 & 9/2 & 11/2\\\hline
g=0 & -13 & 106 & -204 & 111\\
1 & -7 & 118  & -319 & 208\\
2 & -1 & 55 & -219 & 165\\
3 & 0 & 12 & -78 & 66\\ 
4 & 0 & 1 & -14 & 13\\
5 & 0 & 0 & -1 & 1\\\hline
\end{tabular}
                                                                                
$N_{(\lambda^{(1)}+\lambda^{(2)},\,\lambda^{(1)}),g,\beta}^{c=1}$
\end{center}

\begin{center}
\begin{tabular}{|r|cccc|}\hline
\em & $\beta$=5/2 & 7/2 & 9/2 & 11/2\\\hline
g=0 & -22 & 136 & -231 & 117\\
1 & -22 & 231  & -521 & 312\\
2 & -8 & 173 & -532 & 367\\
3 & -1 & 67 & -296 & 230\\
4 & 0 & 13 & -92 & 79\\
5 & 0 & 1 & -15 & 14\\
6 & 0 & 0 & -1 & 1\\\hline
\end{tabular}
                                                                                
$N_{(\lambda^{(3)},\,\lambda^{(1)}),g,\beta}^{c=1}$
\end{center}

$\underline{p_1=2 ~,~  p_2=3}$

\begin{center} 
\begin{tabular}{|r|rr|}\hline
 & $\beta$=2 & 3\\ \hline
g=0 & 1 & -1 \\ \hline
\end{tabular} 

$N_{(\lambda^{(1)},\,\lambda^{(1)}),g,\beta}^{c=1}$
\end{center}  

\begin{center} 
\begin{tabular}{|r|rrr|}\hline
 & $\beta$=5/2 & 7/2 & 9/2 \\ \hline
g=0 & 1 & -5 & 4 \\
1 & 0 & -1 & 1 \\ \hline
\end{tabular} 

$N_{(2\lambda^{(1)},\,\lambda^{(1)}),g,\beta}^{c=1}$
\end{center}  

\begin{center} 
\begin{tabular}{|r|rrr|}\hline
 & $\beta$=5/2 & 7/2 & 9/2 \\ \hline
g=0 & 4 & -13 & 9 \\
1 & 1 & -7 & 6 \\
2 & 0 & -1 & 1 \\ \hline
\end{tabular} 

$N_{(\lambda^{(2)},\,\lambda^{(1)}),g,\beta}^{c=1}$
\end{center}  

\begin{center} 
\begin{tabular}{|r|rrrr|}\hline
 & $\beta$=3 & 4 & 5 & 6 \\ \hline
g=0 & 1 & -15 & 36 & -22 \\
1 & 0 & -7 & 29 & -22 \\
2 & 0 & -1 & 9 & -8 \\
3 & 0 & 0 & 1 & -1 \\ \hline
\end{tabular} 

$N_{(3\lambda^{(1)},\,\lambda^{(1)}),g,\beta}^{c=1}$
\end{center}  

\begin{center} 
\begin{tabular}{|r|rrrr|}\hline
 & $\beta$=3 & 4 & 5 & 6 \\ \hline
g=0 & 13 & -106 & 204 & -111 \\
1 & 7 & -118 & 319 & -208 \\
2 & 1 & -55 & 219 & -165 \\
3 & 0 & -12 & 78 & -66 \\
4 & 0 & -1 & 14 & -13 \\
5 & 0 & 0 & 1 & -1 \\ \hline
\end{tabular} 

$N_{(\lambda^{(1)}+\lambda^{(2)},\,\lambda^{(1)}),g,\beta}^{c=1}$
\end{center}  

\begin{center} 
\begin{tabular}{|r|rrrr|}\hline
 & $\beta$=3 & 4 & 5 & 6 \\ \hline
g=0 & 22 & -136 & 231 & -117 \\
1 & 22 & -231 & 521 & -312 \\
2 & 8 & -173 & 532 & -367 \\
3 & 1 & -67 & 296 & -230 \\
4 & 0 & -13 & 92 & -79 \\
5 & 0 & -1 & 15 & -14 \\
6 & 0 & 0 & 1 & -1 \\ \hline
\end{tabular} 

$N_{(\lambda^{(3)},\,\lambda^{(1)}),g,\beta}^{c=1}$
\end{center}  

\item  Trefoil \# Hopf Link

We consider the link whose one component is trefoil and 
other is Hopf link. We tabulate $N_{R_1,R_2,g,\beta}^{c=1}$ for the case where
both the components carry no framing. 

\begin{center} 
\begin{tabular}{|r|cccc|}\hline
\em & $\beta$=1/2 & 3/2 & 5/2 & 7/2 \\ \hline
g=0 & -3 & 6 & -4 & 1 \\
1 & -1 & 2 & -1 & 0\\ \hline
\end{tabular} 

$N_{(\lambda^{(1)},\,\lambda^{(1)}),g,\beta}^{c=1}$
\end{center}  

\begin{center} 
\begin{tabular}{|r|rrrrrr|}\hline
\em & $\beta$=1 & 2 & 3 & 4 & 5 & 6 \\ \hline  
g=0 & - 14 & 87 & -218 & 266 & -157 & 36 \\ 
1 & -11 & 113 & -415 & 635 & -427 & 105 \\
2 & -2 & 55 & -330 & 650 & -485 & 112 \\
3 & 0 & 12 & -132 & 351 & -285 & 54 \\
4 & 0 & 1 & -26 & 104 & -91 & 12 \\
5 & 0 & 0 & -2 & 16 & -15 & 1 \\
6 & 0 & 0 & 0 & 1 & -1 & 0 \\ \hline
\end{tabular} 

$N_{(2\lambda^{(1)},\,\lambda^{(1)}),g,\beta}^{c=1}$
\end{center}  

\begin{center} 
\begin{tabular}{|r|rrrrrr|}\hline
\em & $\beta$=1 & 2 & 3 & 4 & 5 & 6 \\ \hline
g=0 & -24 & 154 & -363 & 410 & -226 & 49 \\
1 & -26 & 282 & -910 & 1271 & -813 & 196 \\
2 & -9 & 209 & -989 & 1728 & -1233 & 294 \\
3 & -1 & 77 & -572 & 1275 & -989 & 210 \\
4 & 0 & 14 & -182 & 545 & -454 & 77 \\
5 & 0 & 1 & -30 & 135 & -120 & 14 \\
6 & 0 & 0 & -2 & 18 & -17 & 1 \\
7 & 0 & 0 & 0 & 1 & -1 & 0 \\ \hline
\end{tabular} 

$N_{(\lambda^{(2)},\,\lambda^{(1)}),g,\beta}^{c=1}$
\end{center}  
\end{enumerate}

\section{Summary and Discussions}

In this paper, we have briefly presented framed link
invariants in $SO(N)$ Chern-Simons theory. Then, we
studied the expectation value of the observables
in topological string theory carrying $SO$ holonomy.
We had derived modified Frobenius equations leading
to  new polynomial invariants as a reformulation of framed 
link invariants in $SO(N)$ Chern-Simons gauge theory. 
We have proposed new conjectures which are generalisation
of the Ooguri-Vafa conjecture and Bouchard-Florea-Marino
conjecture involving reformulated $SO(N)$ framed link invariants.
We have explicitly computed the reformulated polynomial invariants and 
BPS integer coefficients, corresponding
to cross-cap $c=1$ unoriented topological string
amplitudes, for some non-trivial framed
knots and framed two-component links verifying the conjecture.

It is still a challenging problem of obtaining cross-cap $c=2$ 
unoriented string amplitude on an orientifold of a Calabi-Yau background. 
This requires deriving the amplitude on a covering geometry \cite {vinc2}.

Another open question is to study $SO(N)$ Chern-Simons free-energy at
large $N$ for three-manifolds other than $S^3$. In particular,
we have to pose new duality conjectures involving topological
strings on orientifold background corresponding to $SO(N)$ 
Chern-Simons theory on orbifolds of $S^3$. We hope to study 
these challenging issues in future. 

\newpage
\noindent
{\bf Acknowledgments}  

\noindent
PB would like to thank CSIR for the grant.
The work of PR is supported by Department of Science and Technology
grant under `` SERC FAST TRACK Scheme for Young Scientists''.

\newpage
\appendix{\noindent\bf\Large{Appendix}}

\section{Knot and Link Invariants $V_{\Lambda_{R_1},\Lambda_{R_2},\ldots, 
\Lambda_{R_r}}[L](q,\lambda)$}

In this Appendix we present the knot and link invariants for some knots
and links with arbitrary framings.
\begin{enumerate}
\item {\bf Unknot with framing $p$}

\begin{eqnarray} 
V_{\lambda^{(1)}}&=&(-1)^p\,\lambda^{p/2}\left[1 + \frac{q^{1/2}\,\lambda^{-1/2}\,\left( -1 + \lambda \right) }{-1 + q} \right] \\ \nonumber\\
V_{2\lambda^{(1)}}&=&\frac{q^p\,\lambda^p\,\left( -1 + \lambda \right) \,\left( -q + q^2\,\lambda +
      q^{1/2}\,\lambda^{1/2}\,\left( -1 + q^2 \right)  \right) }{{\left( -1 + q \right) }^2\,\lambda\,
    \left( 1 + q \right) }\\ \nonumber\\
V_{\lambda^{(2)}}&=&\frac{q^{-p}\,\lambda^p\,\left( -1 + \lambda \right) \,\left( -q^2 + q\,\lambda +
      q^{1/2}\,\lambda^{1/2}\,\left( -1 + q^2 \right)  \right) }{{\left( -1 + q \right) }^2\,\lambda\,
    \left( 1 + q \right) }\\ \nonumber \\
V_{3\lambda^{(1)}} & =& \frac{(-1)^p\,q^{3p}\,\lambda^{3p/2}}{{\left( -1 + q \right) }^3\,\left( 1 + q \right) \,\left( 1 + q + q^2 \right) } \left[q\,\lambda^{-3/2}\,\left( -1 + q\,\lambda \right) \,
    \left( q^{1/2} + q^{5/2}\lambda^2 \right.\right.\nonumber\\
&&\left.\left. - q^{1/2}\lambda\,\left( 1 + q^2 \right)  -
      \lambda^{1/2}\,\left( -1 + q^3 \right)  + \lambda^{3/2}\,\left( -1
+ q^3 \right)
      \right)\right]\\ \nonumber\\
V_{\lambda^{(1)}+\lambda^{(2)}}&=&\frac{(-1)^p\lambda^{3p/2}\left( -q + \lambda \right) \left( -1 + q\lambda \right) 
    \left( -q^{3/2} + q^{3/2}\lambda + \lambda^{1/2}\left( -1 + q^3 \right)
      \right) }{{\left( -1 + q \right) }^3\lambda^{3/2}\,\left( 1 + q + q^2 \right) }\\ \nonumber\\
V_{\lambda^{(3)}}&=&\frac{(-1)^p\,q^{-3p}\,\lambda^{3p/2}}{{\left( -1 + q \right) }^3\,
    \left( 1 + q \right) \,\left( 1 + q + q^2 \right) }\left[q\,\lambda^{-3/2}\,\left( -q + \lambda \right) \,
    \left( q^{5/2} + q^{1/2}\,\lambda^2 \right.\right.\nonumber\\
&& \left.\left.-
      q^{1/2}\,\lambda\,\left( 1 + q^2 \right)  - \lambda^{1/2}\,\left( -1
+ q^3 \right)  +
      \lambda^{3/2}\,\left( -1 + q^3 \right)  \right) \right]\\ \nonumber\\
V_{4\lambda^{(1)}}&=&\frac{q^{6p}\lambda^{2p}}{
    {\left( -1 + q \right) }^4\lambda^2{\left( 1 + q \right) }^2\left( 1 +q^2 \right) \,
    \left( 1 + q + q^2 \right) }\left[\left( -1 + \lambda \right) \left( -1 + q\lambda \right)\right. \nonumber \\
&&\left. \left( -1 +
q^2 \lambda\right) 
    \left( -q^2 + q^5\lambda  + q^{3/2}\lambda^{1/2}\left( -1 + q^4 \right)  \right) \right]\\ \nonumber\\
V_{2\lambda^{(1)}+\lambda^{(2)}}&=&\frac{q^{2p}\,\lambda^{2p}}{{\left( -1 + q
\right)
        }^4\,{\left( 1 + q \right) }^2\,\left( 1 + q^2 \right) }\left[q^{1/2}\,\lambda^{-2}\,\left( -q^{1/2} + \lambda^{1/2} \right)
\,
    \left( q^{1/2} + \lambda^{1/2} \right)\right.\nonumber\\
&& \left. \left( q^{3/2} + \lambda^{1/2} \right) \,\left( -1 + \lambda \right) \,
    \left( -1 + q^2\,\lambda \right) \,\left( -1 + q^{5/2}\,\lambda^{1/2} \right) \right]\\ \nonumber\\
V_{2\lambda^{(2)}}&=&\frac{\lambda^{2p}}{{\left( -1 + q \right) }^4\,\lambda^2\,{\left( 1 + q \right) }^2\,
    \left( 1 + q + q^2 \right) }\left[q^4 + q^4\,\lambda^4 -
    q^3\,\lambda^3\,{\left( 1 + q \right) }^2 \right. \nonumber \\
&&   - q^3\,\lambda\,{\left( 1 + q \right) }^2 
 -\left( -1 + q \right) \,q^{5/2}\,\lambda^{1/2}\,{\left( 1 + q \right)
}^2 \nonumber \\
&&   + \left( -1 + q \right) \,q^{5/2}\,\lambda^{7/2}\,{\left( 1 + q \right) }^2 \nonumber \\
&&   + \left( -1 + q \right) \,q^{1/2} 
 \lambda^{3/2}\,{\left( 1 + q \right) }^2
\left( 1 + \left( -1 + q \right) \,q \right) \,\left( 1 + q + q^2 \right) \nonumber \\
&&   - \left( -1 + q \right) \,q^{1/2}\,\lambda^{5/2}\,{\left( 1 + q \right) }^2 
\left( 1 + \left( -1 + q \right) \,q \right) \,\left( 1 + q + q^2 \right)\nonumber \\
&&   - \lambda^2\,\left( 1 + q + q^2 \right) \,
     \left( 1 + q\,\left( -2 + q\,\left( -1 +
             \left( -2 + q \right) \,\left( -1 + q \right) \right.\right.\right.\nonumber \\
&&\left.\left.\left.\left. q\,\left( 1 + q \right)  \right)
          \right)  \right) \right]\\ \nonumber\\
V_{\lambda^{(1)}+\lambda^{(3)}}&=&\frac{q^{-2p}\,\lambda^{2p}}{
    {\left( -1 + q \right) }^4\,{\left( 1 + q \right) }^2\,\left( 1 + q^2 \right) }\left[q^{1/2}\,\lambda^{-2}\,\left( q^{5/2} + \lambda^{1/2} \right) \,
    \left( -1 + \lambda \right)\right. \nonumber\\
&&\left.\left( -q^2 + \lambda \right) \,\left( -1 + q^{1/2}\,\lambda^{1/2} \right) \,
    \left( 1 + q^{1/2}\,\lambda^{1/2} \right) \,\left( -1 + q^{3/2}\,\lambda^{1/2} \right) \right]\\ \nonumber\\
V_{\lambda^{(4)}}&=&\frac{q^{-6p}\lambda^{2p}}{
    {\left( -1 + q \right) }^4\lambda^2{\left( 1 + q \right) }^2\left( 1 +q^2 \right)
    \left( 1 + q + q^2 \right) }\left[\left( -1 + \lambda \right) \left( -q + \lambda \right) \right.\nonumber \\
&&\left. \left( -q^2 + \lambda
\right) 
    \left( -q^5 + q^2\lambda + q^{3/2}\lambda^{1/2}\left( -1 + q^4 \right)  \right) \right]
\end{eqnarray} 

\item {\bf Trefoil knot with framing $p$}

\begin{eqnarray}
V_{\lambda^{(1)}}&=&\frac{(-1)^p}{
    {\left( -1 + q \right) }^2\,\left( 1 + q \right) }\left[q^{-2}\,\lambda^{(p+1)/2}\,\left( q^{3/2} - q^{11/2}
+
      q^6\,\lambda^{1/2} - q^{3/2}\,\lambda \right.\right.\nonumber \\
&& + q\,\lambda^{1/2}
 +
      q^3\,\lambda^{3/2} + q^5\,\lambda^{5/2} -
      q^{5/2}\,\lambda - q^6\,\lambda^{3/2} + q^{9/2}\,\lambda +
      q^{11/2}\,\lambda \nonumber \\
&& - q^3\,\lambda^{1/2} - q^4\,\lambda^{1/2}
 - q\,\lambda^{3/2} +
      q^4\,\lambda^{3/2} + q^{5/2}\,\lambda^2 - q^{9/2}\,\lambda^2 +                                                                                
      q^2\,\lambda^{5/2} \nonumber \\
&&\left.\left. - q^3\,\lambda^{5/2} - q^4\,\lambda^{5/2} \right) \right]\\ \nonumber\\
V_{2\lambda^{(1)}}&=&\frac{1}{{\left( -1 + q \right) }^2\,\left( 1 + q \right) }\left[q^{p-3}\,\lambda^{p+1}\,\left( q^2 + q^5 + q^6 + q^8 +
      {\left( -1 + q \right) }^2 q^9\,\lambda^5 \right.\right.\nonumber \\
&& \left( 1 + q \right)  -
      q^{19/2}\,\lambda^{9/2}\,\left( -1 + q^2 \right)  +
      q^{10}\,\lambda^4\,\left( 1 + q - q^3 \right)  -
      q^2\,\lambda\,\left( 1 + q \right) \nonumber \\
&& \left( 1 + q^2 \right) 
\left( 1 + q^3
+ q^4 \right)  +
      \left( -1 + q \right) \,q^{3/2}\,\lambda^{3/2}\,\left( 1 + q \right) \,
       \left( 1 + \left( -1 + q \right) \,q \right) \nonumber \\
&& \left( 1 + q + q^2 \right)
       \left( 1 + q + q^2 + q^3 + q^4 \right)  -
      \left( -1 + q \right) \,q^{5/2}\,\lambda^{5/2}\,\left( 1 + q \right)\nonumber \\ 
&&       \left( 1 + q^2 \right) 
\left( 1 + q + q^2 + q^3 + q^4 + q^5 + q^6 \right)  +
      q^5\,\lambda^3\,\left( 1 + q \right) \nonumber \\
&& \left( -1 - q^2 - q^4 - q^6 + q^7 \right)  -
      q^{3/2}\,\lambda^{1/2}\,\left( -1 - q^3 + q^7 + q^8 \right) \nonumber \\
&&     + q^{9/2}\,\lambda^{7/2}\,\left( -1 - q^3 + q^7 + q^8 \right)
  -
      q^3\,\lambda^2\,\left( 1 + \left( -1 + q \right) \,q \right)\nonumber \\ 
&&\left.\left.       \left( -1 + q\,\left( -2 + q\,\left( 1 + q^2 \right) 
 \left( -3 - 3\,q + q^3 \right)
            \right)  \right)  \right) \right]\\ \nonumber\\
V_{\lambda^{(2)}}&=&\frac{1}{
    {\left( -1 + q \right) }^2\,\left( 1 + q \right) }\left[q^{-\left(21/2+p\right)}\,\lambda^{p+1}\,
    \left( {\left( -1 + q \right) }^2\,q^{9/2}\,\lambda^5\,\left( 1 + q \right)\right.\right.\nonumber \\
&& - q^5\,\lambda^{9/2}\,\left( -1 + q^2 \right)  +
      q^{7/2}\,\lambda^4\,\left( -1 + q^2 + q^3 \right)  -
      q^{15/2}\,\lambda\,\left( 1 + q \right) \nonumber \\
&& \left( 1 + q^2 \right) \,
       \left( 1 + q + q^4 \right)  + \left( -1 + q \right) q^5\,\lambda^{3/2}\,
       \left( 1 + q \right) \,\left( 1 + \left( -1 + q \right) \,q \right) \nonumber \\
&&      \left( 1 + q + q^2 \right) \,\left( 1 + q + q^2 + q^3 + q^4 \right)  +
      q^{17/2}\,\left( 1 + q^2 + q^3 + q^6 \right) \nonumber \\ 
&&     - \left( -1 + q \right) \,q^4\,\lambda^{5/2}\,\left( 1 + q \right) \,
       \left( 1 + q^2 \right) \left( 1 + q + q^2 + q^3 + q^4 + q^5 + q^6 \right)\nonumber \\
&&     - q^{7/2}\,\lambda^3\,\left( 1 + q \right) \,\left( -1 + q + q^3 + q^5 + q^7 \right)  -
      q^7\,\lambda^{1/2}\,\left( -1 - q + q^5 + q^8 \right) \nonumber \\
&&     + q^4\,\lambda^{7/2}\,\left( -1 - q + q^5 + q^8 \right)  +
      q^{9/2}\,\lambda^2\,\left( -1 +
         q\,\left( 1 + q\,\left( 1 + q\right.\right.\right.\nonumber \\
&&\left.\left.\left.\left.\left. \left( 1 + q + q^2 \right) 
             \left( 1 + q + q^2 + q^4 \right)  \right)  \right)  \right)  \right) \right]
\end{eqnarray}

\item {\bf Hopf Link with framing $p_1$ on first strand and $p_2$ 
on the second}

\begin{eqnarray}
V_{\lambda^{(1)}\,\lambda^{(1)}}&=&\frac{1}{{\left( -1 +
q \right) }^2}\left[{\left( -1 \right) }^{p_1 + p_2}\,
    q^{-1/2}\,\lambda^{\left(p_1+p_2-2\right)/2}\,    \left( q^{1/2} + \lambda^{1/2} \right) \right.\nonumber \\
&& \left( -1 + q^{1/2}\,\lambda^{1/2} \right)
    \left( -1 + q - q^2 + \left( -1 + q \right) \,q^{1/2}\,\lambda^{1/2} \right.\nonumber \\
&&\left.\left. + \lambda\,\left( 1 + \left( -1 + q \right) \,q \right)  \right) \right]\\ \nonumber\\
V_{2\lambda^{(1)}\,\lambda^{(1)}}&=&\frac{1}{{\left( -1 + q \right) }^3\,\left( 1 + q \right) }\left[{\left( -1 \right) }^{p_2}\,
    q^{-1 + p_1}\,\lambda^{\left(2\,p_1+p_2-3\right)/2}\,
    \left( -1 + \lambda^{1/2} \right) \right.\nonumber \\
&& \left( 1 + \lambda^{1/2} \right) 
\left( q^{1/2} + \lambda^{1/2} \right) \,\left( -1 + q^{1/2}\lambda^{1/2} \right) \,
    \left( -1 + q^{3/2}\lambda^{1/2} \right) \nonumber \\
&&\left.    \left( 1 - q + q^3 + q^{1/2}\lambda^{1/2}
\left( 1 + \left( -1 + q \right) \,q^2 \right)
      \right) \right]\\ \nonumber\\
V_{\lambda^{(2)}\,\lambda^{(1)}}&=&\frac{1}{{\left( -1 + q
\right) }^3\,
    \left( 1 + q \right) }\left[{\left( -1 \right) }^{p_2}\,
    q^{-\left(2+p_1\right)}\,\lambda^{\left(2\,p_1+p_2-3\right)/2}\right.\nonumber \\
&&    \left( -\left( q^{7/2}\,\left( 1 + \left( -1 + q \right) \,q^2 \right)  \right)  +
      q^{3/2}\,\lambda^3\,\left( 1 - q + q^3 \right) \right.\nonumber \\
&&     + q^{5/2}\,\lambda\,\left( 1 + q + q^4 \right)  +
      \left( -1 + q \right) \,q^2\,\lambda^{1/2}\,\left( 1 + q + q^4 \right) \nonumber \\
&&     - q^{3/2}\lambda^2\left( 1 + q^3 + q^4 \right)  +
      \left( -1 + q \right) q\lambda^{5/2}\left( 1 + q^3 + q^4 \right)\nonumber \\  
&&\left.\left.     - q\lambda^{3/2}\left( -1 + q^6 \right)  \right) \right]\\ \nonumber\\
V_{3\lambda^{(1)}\,\lambda^{(1)}}&=&\frac{1}{{\left( -1 + q
\right) }^4\,
    \left( 1 + q \right) \,\left( 1 + q + q^2 \right) }\left[{\left( -1 \right) }^{3\,p_1 + p_2}\,
    q^{-3 + 3\,p_1}\,\lambda^{\left(3\,p_1+p_2-4\right)/2}\right.\nonumber \\
&& \left( -1 + \lambda \right) 
\left( q^7\,\lambda^3\,
       \left( 1 + \left( -1 + q \right) \,q^3 \right)  - q^2\,\left( 1 - q + q^4 \right)  \right.\nonumber \\
&&     - q^5\,\lambda^2\,\left( 1 + q + q^5 \right) 
 + q^3\,\lambda\,\left( 1 + q^4 + q^5 \right)  +
      \left( -1 + q \right) \,q^{9/2}\,\lambda^{5/2} \nonumber \\
&& \left( 1 + q + q^2 + q^6 \right)  +
      q^{3/2}\,\lambda^{1/2}\,\left( -1 + q - q^4 + q^7 \right) \nonumber \\
&&\left.\left.     - q^{5/2}\,\lambda^{3/2}\,\left( -1 + q^8 \right)  \right) \right]\\ \nonumber\\
V_{\lambda^{(1)}+\lambda^{(2)}\,\lambda^{(1)}}&=&\frac{1}{{\left(
-1 + q \right) }^4\,
    {\left( 1 + q \right) }^2\,\left( 1 + q + q^2 \right) }\left[{\left(-1\right)}^{3\,p_1+p_2}\,
    q^{-2}\,\lambda^{\left(3\,p_1+p_2-4\right)/2}\right.\nonumber \\
&& \left( q^3 + q^4 + q^8 + q^9 - q^2\,\lambda\,{\left( 1 + q \right) }^2\,\left(
1 + q^2 \right) \,
       \left( 1 + q^4 \right) \right.\nonumber \\
&& - q^2\,\lambda^3\,{\left( 1 + q \right) }^2
\left( 1 + q^2 \right) \,
       \left( 1 + q^4 \right)  + \left( -1 + q \right) \,q^{3/2}\,\lambda^{7/2}\,
       {\left( 1 + q \right) }^2 \nonumber \\
&& \left( 1 + q^2 \right) \,\left( 1 + q^4 \right) 
+ q^2\,\lambda^2\,{\left( 1 + q \right) }^2\,\left( 1 + q + q^2 \right) \,
       \left( 1 + q^4 \right) \nonumber \\
&& + \left( -1 + q \right) \,q^{1/2}\,\lambda^{3/2}\,
       {\left( 1 + q \right) }^2
\left( 1 + q^2 \right) \,\left( 1 + q + q^2 \right) \,
       \left( 1 + q^4 \right) \nonumber \\
&&  - \left( -1 + q \right) \,q^{1/2}\,\lambda^{5/2}\,
       {\left( 1 + q \right) }^2\,\left( 1 + q^2 \right) 
\left( 1 + q + q^2 \right) 
       \left( 1 + q^4 \right) \nonumber \\
&&\left.\left. + q^3\lambda^4\left( 1 + q + q^5 + q^6 \right)
 -
      q^{3/2}\lambda^{1/2}
\left( -1 - q + q^8 + q^9 \right)  \right) \right]
\end{eqnarray}
\begin{eqnarray}
V_{\lambda^{(3)}\,\lambda^{(1)}}&=&\frac{1}{
    {\left( -1 + q \right) }^4\,\left( 1 + q \right) \,\left( 1 + q + q^2 \right) }\left[{\left(-1 \right)}^{3\,p_1 + p_2}\,
    q^{-2 - 3\,p_1}\,\lambda^{\left(-4+3\,p_1+p_2\right)/2}\right.\nonumber\\
&&\left.\left( -1 + \lambda \right) \,\left( -\left( q^6\,
         \left( 1 + \left( -1 + q \right) \,q^3 \right)  \right)  +
      q\,\lambda^3\,\left( 1 - q + q^4 \right) \right.\right.\nonumber\\
&&\left.\left. + q^4\,\lambda\,\left( 1 + q + q^5
\right)  -
      q^2\,\lambda^2\,\left( 1 + q^4 + q^5 \right)  +
      \left( -1 + q \right) \,q^{7/2}\,\lambda^{1/2}\right.\right.\nonumber\\
&&\left.\left.\left( 1 + q + q^2 + q^6 \right)  +
      q^{1/2}\lambda^{5/2}\left( -1 + q - q^4 + q^7 \right) \right.\right.\nonumber \\
&&\left.\left. - q^{3/2}\lambda^{3/2}\left( -1 + q^8 \right)  \right) \right]\nonumber\\ \\
V_{2\lambda^{(1)}\,2\lambda^{(1)}}&=&\frac{1}{{\left( -1 + q \right) }^4\,
    {\left( 1 + q \right) }^2}\left[q^{-4 +p_1+p_2}\,\lambda^{-2+p_1+p_2}\right.\nonumber\\
&& \left.    \left( -\left( q^2\,\lambda\,\left( 1 + q \right) \left( 1 - q + q^3 + q^6 \right)  \right)  -
      \left( -1 + q \right) \,q^{3/2}\,\lambda^{1/2}\,\left( 1 + q \right) \right.\right.\nonumber\\
&&\left.\left. \left( 1 - q + q^3 + q^6 \right)  -
      q^5\,\lambda^3\,\left( 1 + q \right) \,\left( 1 + q^3 - q^5 + q^6 \right)  +                                                                                
      \left( -1 + q \right) \right.\right.\nonumber\\
&&\left.\left.q^{9/2}\,\lambda^{7/2}\,\left( 1 + q \right) \,
       \left( 1 + q^3 - q^5 + q^6 \right)  -
      \left( -1 + q \right) \,q^{5/2}\,\lambda^{5/2}\,\left( 1 + q \right) \right.\right.\nonumber\\
&&\left.\left. \left( 1 + q^2 + q^3 + q^4 + q^5 + q^8 \right)  +
      q^3\,\lambda^2\,\left( 1 + q^3 + 2\,q^4 + q^5 + q^8 \right) \right.\right.\nonumber\\
&&\left.\left. +
      \left( -1 + q \right) \,q^{3/2}\,\lambda^{3/2}\,\left( 1 + q \right) \,
       \left( 1 + q^3 + q^4 + q^5 + q^6 + q^8 \right) \right.\right.\nonumber\\
&&\left.\left. +
      q^6\,\lambda^4\,\left( 1 + \left( -1 + q \right) \,q\,\left( 1 + q \right) \,
          \left( 1 + \left( -1 + q \right) \,q^2 \right)  \right) \right.\right.\nonumber\\
&&\left.\left. +
      q^2\,\left( 1 + \left( -1 + q \right) \,q\,\left( 1 + q \right) \,
          \left( 1 + q - q^2 + q^3 \right)  \right)  \right) \right]\\ \nonumber\\
V_{2\lambda^{(1)}\,\lambda^{(2)}}&=&\frac{1}{
    {\left( -1 + q \right) }^4\,{\left( 1 + q \right) }^2}\left[q^{-2+p_1-p_2}\,\lambda^{-2+p_1+p_2}\,\left( -1 + \lambda \right) \,
    \left( -q^3 + q^5 - q^7 \right.\right.\nonumber\\
&&\left.\left.- \left( -1 + q \right) \,q^{1/2}\,\lambda^{3/2}\,
       {\left( 1 + q \right) }^2\,\left( 1 + \left( -1 + q \right) \,q \right) \,
       \left( 1 + q^4 \right)  + q^3\,\lambda^3\right.\right.\nonumber\\
&&\left.\left.\left( 1 - q^2 +
q^4 \right)  -
      q^2\,\lambda^2\,\left( 1 + q^3 + q^6 \right)  +
      q^2\,\lambda\,\left( 1 + q^3 + q^6 \right) \right.\right.\nonumber\\
&&\left.\left. +
      q^{3/2}\,\lambda^{1/2}\,\left( -1 + q^2 - q^5 + q^7 \right)  +
      \left( -1 + q \right) \,q^{3/2}\,\lambda^{5/2}\,{\left( 1 + q \right) }^2\right.\right.\nonumber\\
&&\left.\left. \left( 1 + \left( -1 + q \right) \,q\,\left( 1 + q^2 \right)  \right)  \right) \right]\\ \nonumber\\
V_{\lambda^{(2)}\,\lambda^{(2)}}&=&\frac{1}{{\left( -1 + q \right) }^4\,{\left( 1 + q \right) }^2}\left[q^{-9/2-p_1-p_2}\,\lambda^{-2+p_1+p_2}\,
    \left( -\left( q^{5/2}\,\lambda^3\,\left( 1 + q \right) \right.\right.\right.\nonumber\\
&&\left.\left.\left. \left( 1 - q + q^3 + q^6 \right)  \right)  +
      \left( -1 + q \right) \,q^2\,\lambda^{7/2}\,\left( 1 + q \right) \,
       \left( 1 - q + q^3 + q^6 \right) \right.\right.\nonumber\\
&&\left.\left. -
      \left( -1 + q \right) \,q^5\,\lambda^{1/2}\,\left( 1 + q \right) \,
       \left( 1 + q^3 - q^5 + q^6 \right)  -
      q^{11/2}\,\lambda\,\left( 1 + q \right) \right.\right.\nonumber\\
&&\left.\left.\left( 1 + q^3 - q^5 + q^6 \right)  +
      \left( -1 + q^2 \right) q^3 \lambda^{3/2} 
       \left( 1 + q^2 + q^3 + q^4 + q^5 + q^8 \right)\right.\right.\nonumber\\
&&\left.\left.  +
      q^{7/2}\,\lambda^2\,\left( 1 + q^3 + 2\,q^4 + q^5 + q^8 \right)  -
      \left( -1 + q \right) \,q^2\,\lambda^{5/2}\,\left( 1 + q \right) \right.\right.\nonumber\\
&&\left.\left. \left( 1 + q^3 + q^4 + q^5 + q^6 + q^8 \right)  +
      q^{13/2}\,\left( 1 + \left( -1 + q \right) \,q\,\left( 1 + q \right) \right.\right.\right.\nonumber\\
&&\left.\left.\left. \left( 1 + \left( -1 + q \right) \,q^2 \right)  \right)  +
      q^{5/2}\lambda^4\,\left( 1 +
         \left( -1 + q \right) q\left( 1 + q \right)\right.\right.\right.\nonumber\\
&&\left.\left.\left. \left( 1 + q - q^2 +
q^3 \right)  \right)
          \right) \right]
\end{eqnarray}
\end{enumerate}

\section{$SO(N)$ Reformulated Invariants $g_{R_1,R_2,\ldots,R_r}(q,\lambda)$}

\begin{enumerate}
\item {\bf Trefoil with framing $p$}

\begin{eqnarray}
g_{\lambda^{(1)}}(q,\lambda)&=&\frac{1}{\left( -1 + q \right) \,q}\left[{\left( -1 \right) }^p\,
    {\lambda }^{\left(1 + p\right)/2}\,
    \left( -\left( {q}^{1/2}\,
         \left( 1 + q^2 \right)  \right) + \left( -1 + q^3 \right) \,{\lambda }^{1/2}\right.\right.\nonumber\\
&& +  {q}^{1/2}\left( 1 + q + q^2 \right) 
       \lambda  - \left( -1 + q^3 \right) 
       {\lambda }^{3/2}  
 - q^{3/2}{\lambda }^2 \nonumber \\
&&\left.\left.     + \left( -1 + q \right) \,q\,
       {\lambda }^{5/2} \right) \right] 
\end{eqnarray} 
\begin{eqnarray} 
g_{2\lambda^{(1)}}(q,\lambda)&=& \frac{1}{2\,{\left( -1 + q \right) }^2\,q^2\,
     \left( 1 + q \right) }\left[2\,{\left( -1 + q \right) }^2\,q^2\,
     \left( 1 + q \right) + {\lambda }^{1 + p} \times \right.\nonumber\\
&&   \left( -2\,q^p\,
        \left( -
             \left( q + q^4 + q^5 + q^7 \right) 
               + {q}^{1/2}\,
           \left( -1 - q^3 + q^7 + q^8 \right){\lambda }^{1/2}\right.\right.\nonumber\\  
&&  +   q\,\left( 1 + q \right) \,
           \left( 1 + q^2 \right) \left( 1 + q^3 + q^4 \right) \,\lambda  -
           \left( -1 + q \right) {q}^{1/2}\,
           \left( 1 + q \right)\nonumber\\
&&  \left( 1 + \left( -1 + q \right) \,q
             \right) \left( 1 + q + q^2 \right)
           \left( 1 + q + q^2 + q^3 + q^4 \right)
           {\lambda }^{3/2}\nonumber\\
&&  + 
          q^2 \left( 1 + 
             \left( -1 + q \right) q \right) 
          \left( -1 + 
             q\,\left( -2 + 
                q\,\left( 1 + q^2 \right) \right.\right.\nonumber \\
&&\left.\left.                 \left( -3 - 3\,q + q^3 \right) 
                \right)  \right) \,{\lambda }^2 + \left( -1 + q \right) q^{3/2}
           \left( 1 + q \right)
           \left( 1 + q^2 \right)\nonumber\\ 
&&   \left( 1 + q + q^2 + q^3 + q^4 + q^5 + 
             q^6 \right) {\lambda }^{5/2}  - q^4\left( 1 + q \right)\nonumber\\
&&    \left( -1 - q^2 - q^4 - q^6 + q^7 \right)
             {\lambda }^3 - 
          q^{7/2}
           \left( -1 - q^3 + q^7 + q^8 \right) 
           {\lambda }^{7/2} \nonumber\\ 
&&   + q^9\,\left( -1 - q + q^3 \right) \,
           {\lambda }^4 + 
          q^{17/2}\,
           \left( -1 + q^2 \right) \,
           {\lambda }^{9/2} \nonumber \\ 
&&\left.         - {\left( -1 + q \right) }^2\,q^8\,
           \left( 1 + q \right) \,{\lambda }^5
          \right)  +  (-1)^p\,
        \left( -\left( (-1)^p \times \right.\right.\nonumber\\  
&&       \left( 1 + q \right) \,
             \left( {q}^{1/2}\,
                  \left( 1 + q^2 \right)  - 
                 \left( -1 + q^3 \right) \,
                  \lambda^{1/2} - q^{1/2}\,
                  \left( 1 + q + q^2 \right)
                  \lambda\right.\nonumber\\ 
&&\left.\left.   + 
                 \left( -1 + q^3 \right) 
                  {\lambda }^{3/2} + 
                 q^{3/2}\,{\lambda }^2 - 
                 \left( -1 + q \right) q
                  {\lambda }^{5/2} \right) 
               ^2 \right) +  \left( -1 + q \right)
           \left( q + \lambda  \right)\nonumber\\  
&&        \left( -1 + q\lambda  \right) 
           \left( -1 + 
             \left( q - \lambda  \right) 
              \left( -\lambda  - q{\lambda }^2  
\left.\left.\left. + q^3
                 \left( -1 + {\lambda }^2 \right) 
                \right)  \right)  \right)  \right) \right]
\end{eqnarray} 
\begin{eqnarray} 
g_{\lambda^{(2)}}(q,\lambda)&=&\frac{1}{2\,{\left( -1 + q \right) }^2\,
    \left( 1 + q \right) }\left[q^{-7 - p}\,{\lambda }^{1 + p}\,
    \left( 2\,\left( q^5 + q^7 + q^8 + q^{11} \right.\right.\right.\nonumber\\
&& - q^{7/2}\,
          \left( -1 - q + q^5 + q^8 \right) \,
          {\lambda}^{1/2} - 
         q^4\left( 1 + q \right) 
          \left( 1 + q^2 \right) \,
          \left( 1 + q + q^4 \right) \lambda \nonumber\\  
&&  + \left( -1 + q \right) \,q^{3/2}\,
          \left( 1 + q \right) \,
          \left( 1 + \left( -1 + q \right) \,q
            \right) \,\left( 1 + q + q^2 \right) \nonumber\\
&&  \left( 1 + q + q^2 + q^3 + q^4 \right) \,
          {\lambda }^{3/2} + 
         q\,\left( -1 + 
            q\,\left( 1 + 
               q\,\left( 1 + 
                  q\,\left( 1 + q + q^2 \right) \right.\right.\right.\nonumber\\
&&\left.\left.\left.  \left( 1 + q + q^2 + q^4 \right) 
                  \right)  \right)  \right) \,
          {\lambda }^2 - 
         \left( -1 + q \right) \,{q}^{1/2}\,
          \left( 1 + q \right) \,
          \left( 1 + q^2 \right) \nonumber\\
&&   \left( 1 + q + q^2 + q^3 + q^4 + q^5 + 
            q^6 \right) {\lambda }^{5/2} -
          \left( -1 + q + q^3 + q^5 + q^7 \right) 
          {\lambda }^3 \nonumber\\ 
&& \left( 1 + q \right) + q^{1/2}
          \left( -1 - q + q^5 + q^8 \right) 
          \lambda^{7/2} + 
         \left( -1 + q^2 + q^3 \right) 
          \lambda^4 \nonumber \\ 
&&        - q^{3/2}\,\left( -1 + q^2 \right) 
          \lambda^{9/2} \left.  + {\left( -1 + q \right) }^2\,q\,
          \left( 1 + q \right) \,{\lambda }^5
         \right)\nonumber\\ 
&&  - {\left( -1 \right) }^p\,
       q^{5 + p}\,\left( {\left( -1 \right) }^p\,
          \left( 1 + q \right) \,
          \left( q^{1/2}\,
               \left( 1 + q^2 \right)  - \left( -1 + q^3 \right) \,
               {\lambda }^{1/2}\right.\right.\nonumber\\   
&&   - q^{1/2}\,
               \left( 1 + q + q^2 \right) \,\lambda 
               + \left( -1 + q^3 \right) \,
               {\lambda }^{3/2} + 
              q^{3/2}\,{\lambda }^2 \nonumber\\  
&&\left.  - \left( -1 + q \right) \,q\,
               {\lambda }^{5/2} \right)^2
          + \left( -1 + q \right) \,
          \left( q + \lambda  \right) \,
          \left( -1 + q\,\lambda  \right) \,
          \left( -1 + 
            \left( q - \lambda  \right) \right.\nonumber\\
&&\left.\left.\left.\left.   \left( -\lambda  - q\,{\lambda }^2 + 
               q^3\,\left( -1 + {\lambda }^2 \right)
                   \right)  \right)  \right) 
      \right) \right]
\end{eqnarray}

\item {\bf Connected sum of trefoil and trefoil with framing $p$}

\begin{eqnarray} 
g_{\lambda^{(1)}}(q,\lambda)&=&\frac{1}{\left( -1 + q \right) \,
    q^2}\left[{\left( -1 \right) }^p {\lambda }^{\left(3 + p\right)/2}
    \left( q^{1/2} +
      {\lambda }^{1/2} \right) \,
    \left( -1 + q^{1/2}\,{\lambda }^{1/2} \right) \right.\nonumber \\
&&\left. {\left( -1 + \left( q^{1/2} - {\lambda }^{1/2}
           \right) \left( q^{1/2}
            \left( q\left( -1 + \lambda  \right)  -
              \lambda  \right)  - {\lambda }^{1/2} \right)
        \right) }^2\right]
\end{eqnarray} 
\begin{eqnarray}
g_{2\lambda^{(1)}}(q,\lambda)&=&\frac{1}{2\,
    {\left( -1 + q \right) }^2\,q^4\,\left( 1 + q \right) }\left[2\,{\left( -1 + q \right) }^2\,q^4\,
     \left( 1 + q \right) \right.\nonumber\\  
&& +  {\lambda }^{3 + p}\left( -\left( 
            {\left( -1 + 
              \left( q^{1/2} - {\lambda }^{1/2} \right)
                 \left( q^{1/2}
                  \left( q\,\left( -1 + \lambda  \right)  - 
                    \lambda  \right)  - {\lambda }^{1/2}
                 \right)  \right) }^4\right.\right.\nonumber\\
&&\left. \left( 1 + q \right) {\left( q^{1/2} + {\lambda }^{1/2} \right) }^
           2{\left( -1 + q^{1/2}{\lambda }^{1/2}
              \right) }^2 \right)  + 
       2\,q^{1/2 + p}
        \left( q^{1/2} + {\lambda }^{1/2} \right) \nonumber \\
&& \left( -1 + q^{3/2}{\lambda }^{1/2}
          \right) \left( -1 + \lambda  \right) 
       \left[ \left( 1 + q^3\left( 1 + q + q^3 \right) \right.\right.\nonumber\\
&&\left.\left.        - \left( -1 + q \right) q^{3/2}
             {\left( 1 + q \right) }^2
             \left( 1 + \left( -1 + q \right) q \right) 
             {\lambda }^{1/2} \right.\right.\nonumber \\ 
&&           - q\,{\left( 1 + q \right) }^2\,
             \left( 1 - q + 2\,q^3 - 2\,q^4 + q^5 \right) \,
             \lambda  + \left( -1 + q \right) \,
             q^{5/2}\,\left( 1 + q \right) \nonumber \\
&&       \left( 1 + q - q^2 + q^3 + q^4 \right) \,
             {\lambda }^{3/2} + 
            q^3\,\left( 1 + 
               \left( -1 + q \right) \,q\,
                \left( 1 + q \right) \right.\nonumber \\
&&\left.     \left( 1 + {\left( -1 + q \right) }^2\,q
                  \right)  \right) \,{\lambda }^2 - 
            q^{9/2}\,
             \left( -1 + q + q^2 - 2\,q^3 + q^5 \right) \,
             {\lambda }^{5/2} \nonumber \\ 
&&\left.\left.          +  {\left( -1 + q \right) }^2\,q^6\,
             \left( 1 + q \right) \,{\lambda }^3 \right) \right]^2
        - {\left( -1 \right) }^p\,\left( -1 + q \right) \,
        \left( q + \lambda  \right)\nonumber \\ 
&&\left.\left.        \left( -1 + q\,\lambda  \right) \,
        {\left( -1 + \left( q - \lambda  \right) \,
             \left( -\lambda  - q\,{\lambda }^2 + 
               q^3\,\left( -1 + {\lambda }^2 \right)  \right)
                \right) }^2 \right) \right] 
\end{eqnarray}
\begin{eqnarray}
g_{\lambda^{(2)}}(q,\lambda)&=&\frac{1}{2\,{\left( -1 + q \right) }^2\,\left( 1 + q \right) }\left[q^{-57/2   - p}\,
    {\lambda }^{3 + p}\,\left( 2\,q^{14}\,
       \left( -1 + {\lambda }^{1/2} \right) \,
       \left( 1 + {\lambda }^{1/2} \right) \right.\right.\nonumber \\
&&  \left( q^{3/2} + {\lambda }^{1/2} \right) \,
       \left( -1 + q^{1/2}\,{\lambda }^{1/2} \right) \,
       \left[ \left( q^{7/2}\,
            \left( 1 + q^2 + q^3 + q^6 \right)  - \left( -1 + q \right)\right.\right.\nonumber \\   
&&          q^3\,
            {\left( 1 + q \right) }^2\,
            \left( 1 + \left( -1 + q \right) \,q \right) \,
            {\lambda }^{1/2} - q^{3/2}\,{\left( 1 + q \right) }^2
          \left( 1 + \left( -1 + q \right) \,q \right)\nonumber \\ 
&&          \left( 1 - q + q^3 \right) \,\lambda   
     + \left( -1 + q \right) \,q\,\left( 1 + q \right) \,
            \left( 1 + q - q^2 + q^3 + q^4 \right) \,
            {\lambda }^{3/2} \nonumber \\ 
&&          + q^{1/2}\,\left( 1 + 
              \left( -1 + q \right) \,q\,
               {\left( 1 + q \right) }^2\,
               \left( 2 + \left( -2 + q \right) \,q \right) 
              \right) \,{\lambda }^2 \nonumber \\
&&\left.\left.      + \left( 1 + q^2\,
               \left( -2 + q + q^2 - q^3 \right)  \right) \,
            {\lambda }^{5/2} + 
           {\left( -1 + q \right) }^2\,q^{1/2}\,
            \left( 1 + q \right) \,{\lambda }^3 \right) \right]^2 \nonumber \\
&&     + q^{49/2 + p}
       \left( -\left(  
            {\left( -1 + 
                \left( q^{1/2} - {\lambda }^{1/2} \right)
                   \left( q^{1/2}
                    \left( q\,
                    \left( -1 + \lambda  \right)  - \lambda 
                     \right)  - {\lambda }^{1/2} \right) 
                \right) }^4\right.\right.\nonumber \\
&&\left.   \left( 1 + q \right) {\left( q^{1/2} + {\lambda }^{1/2} \right) }^
             2\,{\left( -1 + 
                q^{1/2}\,{\lambda }^{1/2} \right) }^2
            \right)  + {\left( -1 \right) }^p\,
          \left( -1 + q \right) \,
          \left( q + \lambda  \right)  \nonumber \\
&&\left.\left.\left.   \left( -1 + q\,\lambda  \right)  {\left( -1 + \left( q - \lambda  \right) \,
               \left( -\lambda  - q\,{\lambda }^2 + 
                 q^3\,\left( -1 + {\lambda }^2 \right) 
                 \right)  \right) }^2 \right)  \right) \right]
\end{eqnarray} 

\item {\bf Hopf Link with framing $p_1$ on first strand and $p_2$ on the second}
                                                       
\begin{eqnarray} 
g_{\lambda^{(1)},\,\lambda^{(1)}}(q,\lambda)&=&{\left( -1 \right) }^
     {p_1 + p_2}\,q^{-1/2}
    \left( q^{1/2} + {\lambda }^{1/2} \right) \,
    \left( -1 + q^{1/2}\,{\lambda }^{1/2} \right) \nonumber \\
&& \left( -1 + \lambda  \right) \,
    {\lambda }^
     {\left(-2 + p_1 + p_2\right)/2} 
\end{eqnarray} 
\begin{eqnarray} 
g_{2\lambda^{(1)},\,\lambda^{(1)}}(q,\lambda)&=&\frac{1}{{\left( -1 + q \right) }^3\,q\,
    \left( 1 + q \right) }\left[{\left( -1 \right) }^{p_2}\,
    \left( -\left(\,
         {\left( -1 + q \right) }^2\,q^{1/2}\,
         \left( 1 + q \right) \right.\right.\right.\nonumber \\
&&\left.  \left( q^{1/2} + {\lambda }^{1/2} \right)
           \,\left( -1 +
           q^{1/2}\,{\lambda }^{1/2} \right)
         \right)  + q^{p_1}\,
       \left( -\left( 
             q^2 + q^{3/2}{\lambda }^{1/2}
                \right) \right.\nonumber \\ 
&&\left.\left. + \left( 1 - q + q^3 +
            q^{1/2}\,
             \left( 1 +
               \left( -1 + q \right) \,q^2 \right) \,
             {\lambda }^{1/2} \right)  \right)\right. \nonumber \\
&&\left.    \left( -1 + q^{3/2}\,
          {\lambda }^{1/2} \right)  \right) \left( q^{1/2} + {\lambda }^{1/2} \right) \,
    \left( -1 + q^{1/2}\,{\lambda }^{1/2} \right)\nonumber \\
&&\left.      \,\left( -1 + \lambda  \right) \,
    {\lambda }^
     {\left(-3 + 2\,p_1 + p_2\right)
       /2}\right]
\end{eqnarray}
\begin{eqnarray}
g_{\lambda^{(2)},\,\lambda^{(1)}}(q,\lambda)&=&\frac{-1}{{\left( -1 + q \right) }^3
      \left( 1 + q \right) }\left[{\left( -1 \right) }^{p_2}
      q^{-1 - p_1}
         \left({\left( -1 + q \right) }^2
         q^{1/2 + p_1}
         \left( 1 + q \right) \right.\right.\nonumber \\
&&       \left( q^{1/2} + {\lambda }^{1/2} \right)
           \,\left( -1 +
           q^{1/2}\,{\lambda }^{1/2} \right)  +
        \left( q^{3/2} + {\lambda }^{1/2}
           \right) \left( -q^{3/2} + \right.\nonumber \\
&&\left.\left.    \left( q^{1/2}
               \left( 1 +
                 \left( -1 + q \right) q^2 \right)  +
               \left( -1 + q - q^3 \right) 
               {\lambda }^{1/2} \right)  +
           q^2\,{\lambda }^{1/2} \right)  \right)\nonumber \\ 
&&\left.      \left( q^{1/2} + {\lambda }^{1/2} \right) \,
      \left( -1 + q^{1/2}\,{\lambda }^{1/2} \right)
        \,\left( -1 + \lambda  \right) \,
      {\lambda }^
       {\left(-3 + 2\,p_1 + p_2\right)
         /2}\right] 
\end{eqnarray} 
\begin{eqnarray} 
g_{3\lambda^{(1)},\,\lambda^{(1)}}(q,\lambda)&=&\frac{1}{{\left( -1 + q \right) }^2\,
    q^{5/2}\,\left( 1 + q \right) }\left[{\left( -1 \right) }^
     {p_1 + p_2}\,
    \left( q^{1/2} + {\lambda }^{1/2} \right) \,
    \left( -1 + q^{1/2}\,{\lambda }^{1/2} \right)\right.\nonumber \\
&&    \,\left( -1 + \lambda  \right) \,
    {\lambda }^
     {\left(-4 + 3\,p_1 + p_2\right)
       /2}\left( q^2\left( 1 + q \right) 
       {\left( q^{1/2} + {\lambda }^{1/2} \right)
           }^2 {\left( -1 +
           q^{1/2}{\lambda }^{1/2} \right) }^2 \right.\nonumber \\
&&     + q^{1 + 3\,p_1}
       \left( 1 + q^{1/2}{\lambda }^{1/2} \right)
         \left( -1 + q\,{\lambda }^{1/2} \right) 
       \left( 1 + q\,{\lambda }^{1/2} \right) 
       \left( -1 + q^{5/2}
          {\lambda }^{1/2} \right)\nonumber \\  
&&\left.\left.     - q^{3/2 + p_1}
       \left( q^{1/2} + {\lambda }^{1/2} \right) 
       \left( -1 + q^{3/2}
          {\lambda }^{1/2} \right) 
       \left( -1 - 2\,q +
         q\,\left( 2 + q \right) \lambda  \right)
      \right) \right]\nonumber \\ 
\end{eqnarray} 
\begin{eqnarray} 
g_{\lambda^{(1)}+\lambda^{(2)},\,\lambda^{(1)}}(q,\lambda)&=&\frac{-1}{{\left( -1 + q \right) }^2\,
      \left( 1 + q \right) }\left[{\left( -1 \right) }^
       {p_1 + p_2}\,
      q^{-3/2   -
         p_1}\,
      \left( q^{1/2} + {\lambda }^{1/2} \right) \right.\nonumber \\
&&     \left( -1 + q^{1/2}{\lambda }^{1/2} \right)
        \left( -1 + \lambda  \right) 
      {\lambda }^
       {\left(-4 + 3\,p_1 + p_2\right)
         /2}\left( -\left( q^{1/2}
           \left( q^{3/2} +
             {\lambda }^{1/2} \right) \right.\right.\nonumber \\
&&\left.           \left( -1 +
             q^{1/2}{\lambda }^{1/2} \right) 
           \left( q\left( 2 + q - 2\,\lambda  \right)
                 - \lambda  \right)  \right)  +
        q^{1/2 + 2\,p_1}
         \left( q^{1/2} + {\lambda }^{1/2} \right)\nonumber \\
&&           \left( -1 +
           q^{3/2}\,{\lambda }^{1/2} \right) 
          \left( -1 - 2\,q +
           q\,\left( 2 + q \right) \,\lambda  \right)
         + q^{p_1}\,\left( 1 + q \right)\nonumber \\ 
&&   \left( -3\,q^2 +
           \left( -1 + q \right) \,q^{1/2}\,
            \left( 1 + q\,\left( 4 + q \right)  \right)
              \,{\lambda }^{1/2} \right.\nonumber \\
&& + \left( 1 + q\,
               \left( -3 +
                 q \left( 10
   +   \left( -3 + q \right) q \right)
                 \right)  \right) \lambda \nonumber \\ 
&&\left.\left.\left.          - \left( -1 + q \right) q^{1/2}
            \left( 1 + q\left( 4 + q \right)  \right)
              {\lambda }^{3/2} -
           3q^2{\lambda }^2 \right)  \right) \right] 
\end{eqnarray} 
\begin{eqnarray} 
g_{\lambda^{(3)},\,\lambda^{(1)}}(q,\lambda)&=&\frac{1}{{\left( -1 + q \right) }^2\,\left( 1 + q \right) }\left[{\left( -1 \right) }^
     {p1 + p2}\,
    q^{-3/2   -
       3\,p1}
    \left( -1 + q^{1/2}{\lambda }^{1/2} \right)
      \left( -1 + \lambda  \right) \right.\nonumber \\
&&    {\lambda }^
     {\left(-4 + 3\,p1 + p2\right)
       /2}\,\left( q^{1 + 3\,p1}\,
       \left( 1 + q \right) \,
       {\left( q^{1/2} + {\lambda }^{1/2} \right)
           }^3\,{\left( -1 +
           q^{1/2}\,{\lambda }^{1/2} \right) }^2 \right.\nonumber \\
&&     + q^{1/2 + 2\,p1}\,
       \left( q^{1/2} + {\lambda }^{1/2} \right) \,
       \left( q^{3/2} + {\lambda }^{1/2}
         \right) \,\left( -1 +
         q^{1/2}\,{\lambda }^{1/2} \right) \nonumber \\
&&\left.\left.       \left( q\left( 2 + q - 2\,\lambda  \right)  -
         \lambda  \right)  +
      \left( q^2 - \lambda  \right) 
       \left( q^{7/2} + q\,{\lambda }^{1/2} -
         q^{5/2}\lambda  -
         {\lambda }^{3/2} \right)  \right) \right]
\end{eqnarray} 
\begin{eqnarray} 
g_{2\lambda^{(1)},\,2\lambda^{(1)}}(q,\lambda)&=&
\frac{1}{2\,{\left( -1 + q \right) }^4 q\,\left( 1 + q \right)}\left[{\lambda }^
     {-2 + p1 + p2}\,
    \left( {\left( q^{1/2} +
             {\lambda }^{1/2} \right) }^2\,
         {\left( -1 +
             q^{1/2}\,{\lambda }^{1/2} \right) }^2 \right.\right.\nonumber \\
&&    \left( 2\,q^{3/2 + p2}\,
            \left( q^{1/2} + {\lambda }^{1/2}
              \right) \,
            \left( -1 +
              q^{3/2}\,{\lambda }^{1/2}
              \right) \,\left( -1 + \lambda  \right)  +
            2\,q^{p1}\,
            \left( -1 + \lambda  \right)\right.\nonumber \\ 
&&           \left( -q^2 +
              q^{3/2}\,
               \left( -1 + q^2 \right) \,
               {\lambda }^{1/2} + q^3\,\lambda
              \right)  -
           \left( 1 + q \right) \,
            \left( -1 - \left( -5 + q \right) \,q \right.\nonumber \\
&&\left.        - 3\,\left( -1 + q \right) \,q^{1/2}\,
               {\lambda }^{1/2} +
              \left( 1 + \left( -5 + q \right) \,q
                 \right) \,\lambda  \right) 
            \left( -1 + q - q^2 \right.\nonumber \\
&&\left.\left.            + \left( -1 + q \right) \,q^{1/2}\,
               {\lambda }^{1/2} +
              \left( 1 + \left( -1 + q \right) \,q
                 \right) \,\lambda  \right)  \right)  
     + {\left( -1 \right) }^
{p1 + p2}\,\left( q+1 \right) \nonumber \\
&&         \left( -3\,{\left( -1 \right) }^
             {p1 + p2}\,q\,
            {\left( q^{1/2} + {\lambda }^{1/2}
                \right) }^4\,
            {\left( -1 +
                q^{1/2}\,{\lambda }^{1/2} \right) }
              ^4 
  - {\left( -1 + q \right) }^4\,
            \left( q + \lambda  \right) \right.\nonumber \\
&&\left.            \left( -1 + q\,\lambda  \right) \,
            \left( -1 + {\lambda }^2 \right)  \right) 
          + 2\,\left( -1 + {\lambda }^{1/2} \right)
         \,\left( 1 + {\lambda }^{1/2} \right) 
       \left( q^{1/2} + {\lambda }^{1/2} \right) \nonumber \\
&&       \left( -1 + q^{1/2}\,{\lambda }^{1/2}
         \right) \,\left( -1 +
         q^{3/2}\,{\lambda }^{1/2} \right) \,
       \left( - \left( q^
                 {p1} + q^{p2}
                \right) 
       \left( q^{1/2} +
                {\lambda }^{1/2} \right) \right.\nonumber \\
&&              \left( -1 +
                q^{1/2}\,{\lambda }^{1/2} \right)
               \left( 1 - q + q^3 +
                q^{1/2}\,
                 \left( 1 +
                   \left( -1 + q \right) \,q^2 \right)
                  \,{\lambda }^{1/2} \right) \nonumber \\  
&&        + q\,\left(q+1\right){\left( -1 + q \right) }^2\,
          q^
           {-\left( 5/2 \right)  +
             p1 + p2}\,
          \left( 1 + q^{1/2}\,{\lambda }^{1/2}
\right) \nonumber \\
&&\left.\left.\left.           \left( -1 +
            q\,\left( 1 - q +
               \left( -1 + q \right) \,q^{1/2}\,
                {\lambda }^{1/2} 
              + q\,\left( 1 +
                  \left( -1 + q \right) \,q \right) \,
                \lambda  \right)  \right)  \right)
      \right) \right]\nonumber \\ 
\end{eqnarray} 
\begin{eqnarray} 
g_{2\lambda^{(1)},\,\lambda^{(2)}}(q,\lambda)&=&
\frac{1}{2\,
    {\left( -1 + q \right) }^4 q\, \left( q+1 \right)} \left[{\lambda }^
     {-2 + p1 + p2}\,
    \left( {\left( -1 \right) }^
          {p1 + p2}\, \left( q+1 \right)
         \left( -3\,{\left( -1 \right) }^
             {p1 + p2}\right.\right.\right.\nonumber \\
&&           q {\left( q^{1/2} + {\lambda }^{1/2}
                \right) }^4
            {\left( -1 + 
                q^{1/2}{\lambda }^{1/2} \right) }
              ^4 + {\left( -1 + q \right) }^4
            \left( q + \lambda  \right) 
            \left( -1 + q\lambda  \right) \nonumber \\
&&\left.            \left( -1 + {\lambda }^2 \right)  \right) 
          + {\left( q^{1/2} + 
           {\lambda }^{1/2} \right) }^2\,
       {\left( -1 + q^{1/2}\,{\lambda }^{1/2}
           \right) }^2\,
       \left( - \left( q+1 \right) \right.\nonumber \\
&&  \left( -1 - 
                \left( -5 + q \right) \,q 
             - 3\left( -1 + q \right) q^{1/2}
                 {\lambda }^{1/2} + 
                \left( 1 + 
                  \left( -5 + q \right) q \right) 
                 \lambda  \right) \nonumber \\
&&              \left( -1 + q - q^2 + 
                \left( -1 + q \right) 
 q^{1/2}
                 {\lambda }^{1/2} + 
                \left( 1 + 
                  \left( -1 + q \right) \,q \right) \,
                 \lambda  \right) \nonumber \\
&&        + 2\,q\,\left( -1 + \lambda  \right) \,
            \left( q^{1/2 - p2}\,
               \left( q^{3/2} + 
                 {\lambda }^{1/2} \right) 
             \left( -1 + 
                 q^{1/2}\,{\lambda }^{1/2} \right)\right.\nonumber \\ 
&&\left.\left.               + q^{p1}\,
               \left( -q + 
                 q^{1/2}\,\left( -1 + q^2 \right) \,
                  {\lambda }^{1/2} + q^2\,\lambda 
                 \right)  \right)  \right) 
     + 2\,q^{-\left( 3/2 \right)  - 
          p2} \nonumber \\
&&       \left( q^{1/2} + {\lambda }^{1/2} \right) \,
       \left( -1 + q^{1/2}\,{\lambda }^{1/2}
         \right) \,\left( -  q\, \left( q+1 \right) {\left( -1 + q
                \right) }^2\,q^{p1} \right.\nonumber \\
&&            \left( q^{1/2} - {\lambda }^{1/2}
              \right) 
            \left( -1 + {\lambda }^{1/2} \right) \,
            \left( 1 + {\lambda }^{1/2} \right) \,
            \left( q^{3/2} + 
              {\lambda }^{1/2} \right) \,
            \left( 1 + 
              q^{1/2}\,{\lambda }^{1/2} \right) \nonumber \\
&&            \left( -1 + 
              q^{3/2}\,{\lambda }^{1/2}
              \right)   
        - q^{3/2}\,
            \left( q^{1/2} + {\lambda }^{1/2}
              \right) \,
            \left( -1 + 
              q^{1/2}\,{\lambda }^{1/2} \right) \,
            \left( -1 + \lambda  \right) \nonumber \\
&&            \left( -q^2 + q^4 - q^5 
              + {\lambda }^{1/2} \left( -
              q^{1/2} + 
              q^{3/2} - 
              q^{7/2} + 
              q^{9/2} \right) + 
              \lambda \left(1 - q  + q^3 \right) \right.\nonumber \\  
&&       + q^{p1 + p2}\,
               \left( -1 + q - q^3 + 
                 q^{1/2}\,
                  \left( -1 + q - q^3 + q^4 \right) \,
                  {\lambda }^{1/2} \right.\nonumber \\
&&\left.\left.\left.\left.\left.                + q^2\,\left( 1 + 
                   \left( -1 + q \right) \,q^2 \right) 
                   \,\lambda  \right)  \right) 
         \right)  \right) \right]
\end{eqnarray} 
\begin{eqnarray} 
g_{\lambda^{(2)},\,\lambda^{(2)}}(q,\lambda)&=&\frac{1}{2\,
    {\left( -1 + q \right) }^4\,\left( 1 + q \right) }\left[q^{-\left( 9/2 \right)  - 
       p1 - p2}\,
    {\lambda }^{-2 + p1 + p2}\,
    \left( 2\,{\left( -1 + q \right) }^2\,q^2\,
       \left( 1 + q \right) \right.\right.\nonumber \\
&&       \left( q^{1/2} - {\lambda }^{1/2} \right) \,
       \left( -1 + {\lambda }^{1/2} \right) \,
       \left( 1 + {\lambda }^{1/2} \right) \,
       \left( q^{1/2} + {\lambda }^{1/2} \right) \,
       \left( q^{3/2} + {\lambda }^{1/2}
         \right) \nonumber \\
&&  \left( -1 + 
         q^{1/2}\,{\lambda }^{1/2} \right) 
       \left( q^2 - q^3 + q^4 + 
         q^{3/2 + p1}
          \left( q^{1/2} + {\lambda }^{1/2} \right)
            \left( -1 + 
            q^{1/2}{\lambda }^{1/2} \right) \right.\nonumber \\
&&\left.        + q^{3/2}\,{\lambda }^{1/2} - 
         q^{5/2}\,{\lambda }^{1/2} - 
         \lambda  + q\,\lambda  - q^2\,\lambda  \right)
          + q^{p2}\,\left( 1 + q \right) \,
       \left( 2\,{\left( -1 + q \right) }^2 \right.\nonumber \\
&&          q^{7/2}\,
          {\left( q^{1/2} + {\lambda }^{1/2}
              \right) }^2\,
          {\left( -1 + 
              q^{1/2}\,{\lambda }^{1/2} \right) }^2
           \,\left( q^2 - 
            \left( -1 + q \right) \,q^{1/2}\,
             {\lambda }^{1/2} - \lambda  \right) \nonumber \\
&&          \left( -1 + \lambda  \right)  - 
         q^{7/2 + p1}\,
          \left( -1 + 
            q^{1/2}\,{\lambda }^{1/2} \right) \,
          \left( 3\,{\left( -1 \right) }^
              {2\,\left( p1 + 
                  p2 \right) }\,q\,
             {\left( q^{1/2} + 
                 {\lambda }^{1/2} \right) }^4 \right.\nonumber \\
&&             {\left( -1 + 
                 q^{1/2}\,{\lambda }^{1/2} \right) 
                }^3 + {\left( q^{1/2} + 
                 {\lambda }^{1/2} \right) }^2\,
             \left( -1 + 
               q^{1/2}\,{\lambda }^{1/2} \right) \,
             \left( -1 - \left( -5 + q \right) \,q \right.\nonumber \\ 
&&\left.     - 3\,\left( -1 + q \right) \,q^{1/2}\,
                {\lambda }^{1/2} + 
               \left( 1 + 
                 \left( -5 + q \right) \,q \right) \,
                \lambda  \right) \,
             \left( -1 + q - q^2 + 
               \left( -1 + q \right) \right.\nonumber \\
&&\left.  q^{1/2}\,
                {\lambda }^{1/2} + 
               \left( 1 + 
                 \left( -1 + q \right) \,q \right) \,
                \lambda  \right)  + 
            {\left( -1 \right) }^
              {p1 + p2}\,
             {\left( -1 + q \right) }^4\,
             \left( 1 + 
               q^{1/2}\,{\lambda }^{1/2} \right) \nonumber \\
&&\left.\left.\left.\left.             \left( q + \lambda  \right)\, 
             \left( -1 + {\lambda }^2 \right)  \right) 
         \right)  \right) \right]
\end{eqnarray} 

\item {\bf Connected sum of Trefoil and Hopf link
with framing $p_1$ on Trefoil and framings $p_1$ and $p_2$ on the hopf link}

\begin{eqnarray} 
g_{\lambda^{(1)},\,\lambda^{(1)}}(q,\lambda)&=& q^{-3/2}\left[{\left( -1 \right) }^
     {p_1 + p_2}\,
    \left( -1 + \lambda  \right) \,
    {\lambda }^
     {\left(p_1 + p_2\right)/2}\,
    \left( -\left( q^{1/2}\,
         \left( 1 + q^2 \right)  \right) \right.\right.\nonumber \\ 
&&     + \left( -1 + q^3 \right) \,{\lambda }^{1/2} +
      q^{1/2}\,\left( 1 + q + q^2 \right) \,
       \lambda  - \left( -1 + q^3 \right) \,
       {\lambda }^{3/2} \nonumber \\
&&\left.\left.     - q^{3/2}\,{\lambda }^2 +
      \left( -1 + q \right) \,q\,{\lambda }^{5/2}
      \right) \right]\\ \nonumber \\
g_{2\lambda^{(1)},\,\lambda^{(1)}}(q,\lambda)&=&\frac{-1}{\left( -1 + q \right) \,q^3}\left[{\left( -1 \right) }^{p_2}\,
      \left( -1 + {\lambda }^{1/2} \right) \,
      \left( 1 + {\lambda }^{1/2} \right) \,
      \left( q^{1/2} + {\lambda }^{1/2} \right) \right.\nonumber \\
&&     \left( -1 + q^{1/2}\,{\lambda }^{1/2} \right)
        \,{\lambda }^
       {\left(1 + 2\,p_1 + p_2\right)/2}
       \,\left( q^{1/2}\,
         \left[\left( -1 + \left( q^{1/2} -
                {\lambda }^{1/2} \right) \right.\right.\right.\nonumber \\
&&\left.\left.    \left( q^{1/2}
                 \left( q
                   \left( -1 + \lambda  \right)  -
                   \lambda  \right)  - {\lambda }^{1/2}
                \right)  \right) \right]^2
         \left( q^{1/2} + {\lambda }^{1/2} \right) 
         \left( -1 + q^{1/2}\,{\lambda }^{1/2}
           \right) \nonumber \\
&&  + q^{p_1}\,
         \left( 1 + q^3 + q^4 + q^6 +
           q^{1/2}\,
            \left( 1 -
              q^4\,\left( -1 + q + q^3 \right)  \right) 
             {\lambda }^{1/2} \right.\nonumber \\
&&  - q\,\left( 1 + q + 2\,q^3 + q^4 + q^5 + q^6 +
              q^7 \right) \,\lambda  \nonumber \\
&&          + q^{3/2}\,
            \left( -1 - q + q^7 + q^8 \right) \,
            {\lambda }^{3/2} +
           q^3\,\left( 1 + q^2 + 2\,q^3 + q^6 \right) \,
            {\lambda }^2 \nonumber \\
&&          + q^{7/2}\,
            \left( 1 + q^3 - q^4 + q^6 - 2\,q^7 \right)
              \,{\lambda }^{5/2} +
           q^8\,\left( -1 +
              \left( -1 + q \right) \,q \right) \,
            {\lambda }^3 \nonumber \\
&&\left.\left.\left.          + q^{15/2}\,
            \left( -1 + q - q^3 + q^4 \right) \,
            {\lambda }^{7/2} -
           {\left( -1 + q \right) }^2\,q^8\,
            \left( 1 + q \right) \,{\lambda }^4 \right)
        \right) \right] \\ \nonumber \\
g_{\lambda^{(2)},\,\lambda^{(1)}}(q,\lambda)&=& 
 \frac{-1}{-1 + q}\left[{\left( -1 \right) }^{p_2}\,
      q^{-17/2   - 
         p_1}\,
      \left( q^{1/2} + {\lambda }^{1/2} \right) 
      \left( -1 + q^{1/2}\,{\lambda }^{1/2} \right)
        \left( -1 + \lambda  \right) \right.\nonumber \\
&&      {\lambda }^
       {\left(1 + 2\,p_1 + p_2\right)/2}
       \,\left( q^{11/2} + q^{15/2} + 
        q^{17/2} + q^{23/2} + 
        q^{6 + p_1}\right.\nonumber \\
&&         {\left( -1 + \left( q^{1/2} - 
                {\lambda }^{1/2} \right) \,
              \left( q^{1/2}\,
                 \left( q\,
                   \left( -1 + \lambda  \right)  - 
                   \lambda  \right)  - {\lambda }^{1/2}
                \right)  \right) }^2 \nonumber \\
&&         \left( q^{1/2} + {\lambda }^{1/2} \right) \,
         \left( -1 + q^{1/2}\,{\lambda }^{1/2}
           \right)  + {\lambda }^{1/2} \left(   q^4 + 
        q^6 - 
        q^7 - 
        q^{11} \right)  \nonumber \\
&&     + \lambda \left(  - q^{7/2}  - 
        q^{9/2}  - 
        q^{11/2}  - 
        q^{13/2}  - 
        2\,q^{15/2}  - 
        q^{19/2}  - 
        q^{21/2} \right)  \nonumber \\ 
&&  +{\lambda }^{3/2} \left( - q^2 - 
        q^3 + 
        q^9 + 
        q^{10} \right) + 
        {\lambda }^2 \left( q^{5/2} + 
        2\,q^{11/2} + 
        q^{13/2} + 
        q^{17/2} \right)  \nonumber \\
&&       + {\lambda }^{5/2} \left( 2\,q - 
        q^2 + 
        q^4 - 
        q^5 - 
        q^8 \right) + {\lambda }^3 \left( 
        q^{3/2} - 
        q^{5/2} - 
        q^{7/2} \right)  \nonumber \\
&&\left.\left.  + {\lambda }^{7/2} \left( -1 + 
        q - 
        q^3 + 
        q^4 \right) + {\lambda }^4 \left( 
        -q^{1/2} + 
        q^{3/2} + 
        q^{5/2} - 
        q^{7/2}\right) \right) \right]
\end{eqnarray} 
\end{enumerate}


\begin{thebibliography}{99}
\bibitem{wittencs}
E. Witten, ``Chern-Simons Gauge Theory as a String Theory,''
{\tt hep-th/9207094}
\bibitem{gv1}
R. Gopakumar, C. Vafa, ``M-Theory and Topological Strings, I,''
{\tt hep-th/9809187}.
\bibitem{gv2}
R. Gopakumar, C. Vafa, ``On the Gauge Theory/ Geometry Correspondence,''
{\tt hep-th/9811131}.
\bibitem{gv3}
R. Gopakumar, C. Vafa, ``M-Theory and Topological Strings, II,''
hep-th/9812127.
\bibitem{ov}
H. Ooguri, C. Vafa, ``Knot Invariants and Topological Strings,''
Nucl. Phys. {\bf B577}, 419, (2000), {\tt hep-th/9912123}.
\bibitem{lm}
J. M. F. Labastida, M. Marino, ``Polynomial Invariants for Torus Knots
and Topological Strings,'' {\tt hep-th/0004196}.
\bibitem{taps}
P. Ramadevi, T. Sarkar, ``On Link Invariants and Topological String
Amplitudes,'' Nucl. Phys. {\bf B 600} (2001) 487.
\bibitem{laba}
J. M. F Labastida, M. Marino, C. Vafa, ``Knots, Links and Branes
at Large N,'' JHEP11 (2000) 007, {\tt hep-th/0010102}
\bibitem{mari} M. Marino, C. Vafa, `` Framed Knots at Large $N$,'' 
{\tt hep-th/0108064}.
\bibitem{ramprav} Pravina Borhade, P. Ramadevi, Tapobrata Sarkar,
`` U(N) Framed Links, Three-Manifold Invariants, and Topological Strings,''
Nucl.Phys. B678 (2004) 656-681.
\bibitem{thoo} 
G. `t Hooft, ``A Planar Diagram Theory for Strong Interactions,''
Nucl. Phys. {\bf B 72} (1974) 461.
\bibitem{marin}M. Marino, `` Chern-Simons theory, matrix integrals and
perturbative three-manifold invariants,'' Commun.Math.Phys. 253 (2004) 25-49.
\bibitem{dijk} R. Dijkgraaf and C. Vafa, `` Matrix models,
topological strings and supersymmetric gauge theories,'' Nucl. Phys.
{\bf B 644} 3 (2002).
\bibitem{akmv} M. Aganagic, A. Klemm, M. Marino and C. Vafa, 
``Matrix model as a mirror of Chern-Simons theory,'' JHEP 0402 (2004) 010.
\bibitem{sinha}S. Sinha, C. Vafa, ``SO and Sp Chern-Simons at Large N,''
{\tt hep-th/0012136}.
\bibitem{vinc1}V. Bouchard, B. Florea, M. Marino,``Counting Higher Genus 
Curves with Crosscaps in Calabi-Yau Orientifolds,'' JHEP 0412(2004) 035, 
{\tt hep-th/0405083}
\bibitem{vinc2}V. Bouchard, B. Florea, M. Marino,
``Topological Open String Amplitudes On Orientifolds,'' JHEP 0502 (2005) 002,
{\tt hep-th/0411227}.
\bibitem{labb} J.M.F. Labastida, M. Marino, `` A New Point of
View in the Theory of Knot and Link Invariants,'' {\tt Math.QA/0104180},
J. Knot Theory Ramifications {\bf 11} (2002) 173.
\end{thebibliography}
\end{document}